\documentclass[twocolumn,tighten, table]{aastex631}

\usepackage{natbib}
\usepackage{amsmath}
\usepackage{listings}
\usepackage[normalem]{ulem}
\usepackage[section]{placeins}
\usepackage{color}
\usepackage{graphicx}
\usepackage{mathtools}
\usepackage{multirow}
\usepackage{xcolor}

\usepackage{fancyvrb}

\shorttitle{Revisiting dynamical friction: modes and wakes}
\shortauthors{T.~Tamfal et al.}

\begin{document}

\title{{\Large \bf Revisiting dynamical friction: the role of global modes and local wakes}}


\correspondingauthor{Tomas Tamfal}
\email{tomas.tamfal@uzh.ch}

\author[0000-0003-1773-9349]{Tomas Tamfal}
\affiliation{Center for Theoretical Astrophysics and Cosmology, Institute for Computational Science, University of Zurich, Winterthurerstrasse 190, CH-8057 Z\"urich, Switzerland}

\author[0000-0002-7078-2074]{Lucio Mayer}
\affiliation{Center for Theoretical Astrophysics and Cosmology, Institute for Computational Science, University of Zurich, Winterthurerstrasse 190, CH-8057 Z\"urich, Switzerland}

\author{Thomas R. Quinn}
\affiliation{Department of Astronomy, University of Washington, Seattle, WA 98195-1580, USA}

\author[0000-0002-1786-963X]{Pedro R. Capelo}
\affiliation{Center for Theoretical Astrophysics and Cosmology, Institute for Computational Science, University of Zurich, Winterthurerstrasse 190, CH-8057 Z\"urich, Switzerland}

\author{Stelios Kazantzidis}
\affiliation{Section of Astrophysics, Astronomy and Mechanics, Department of Physics, National and Kapodistrian University of Athens, 15784 Zografos, Athens, Greece}

\author[0000-0003-1746-9529]{Arif Babul}
\affiliation{Department of Physics and Astronomy, University of Victoria, Victoria, BC V8P 1A1, Canada}

\author[0000-0002-0757-5195]{Douglas Potter}
\affiliation{Center for Theoretical Astrophysics and Cosmology, Institute for Computational Science, University of Zurich, Winterthurerstrasse 190, CH-8057 Z\"urich, Switzerland}


\begin{abstract}
The orbital decay of a perturber within a larger system plays a key role in the dynamics of many astrophysical systems -- from nuclear star clusters or globular clusters in galaxies, to massive black holes in galactic nuclei, to dwarf galaxy satellites within the dark matter halos of more massive galaxies. For many decades, there have been various attempts to determine the underlying physics and time-scales of the drag mechanism, ranging from the local dynamical friction approach to descriptions based on the back-reaction of global modes induced in the background system. We present ultra-high-resolution $N$-body simulations of massive satellites orbiting a Milky Way-like galaxy (with $> 10^8$ particles), that appear to capture both the local \textit{``wake''} and the global \textit{``mode''} induced in the primary halo. We address directly the mechanism of orbital decay from the combined action of local and global perturbations and specifically analyze where the bulk of the torque originates.
\end{abstract}



\section{Introduction}\label{sec:Introduction}
In the modern theory of structure formation based on cold dark matter (CDM), galaxies form in dark matter (DM) halos whose assembly is the result of repeated mergers with other halos \citep[][]{Blumenthal_et_al_1984}. Minor mergers, in which a primary halo and a much lighter secondary halo interact and eventually merge, are the most common modes of structure formation \citep[][]{Fakhouri_et_al_2010,OLeary_et_al_2020}. In this case, the secondary halo orbits inside the more massive halo, after being captured by its gravity, and gradually sinks to the center due to a gravitational drag force exerted by the primary halo, known as ``dynamical friction'' \citep[][]{Chandrasekhar_1943}. Since the earliest attempts to model galaxy formation in the CDM scenario, which involved using semi-analytical models, dynamical friction has played the role of one of the key physical processes necessary to understand the cosmic evolution of structures \citep[][]{Lacey_et_al_1993, Taylor_et_al_2001, Benson_et_al_2002}.

Yet the study of the theoretical foundations of dynamical friction pre-dates these works, and has since the beginning led to different approaches, and often to contentions between such approaches which are not yet settled. One explanation is that dynamical friction is a drag force arising from a global response of the primary halo to the perturbation induced by the satellite. This description stands in contrast to the popular interpretation of the original Chandrasekhar's formula \citep[][]{Chandrasekhar_1943} of dynamical friction being a local drag caused by the overdensity trailing the perturber \citep[][]{Mulder_et_al_1983,Colpi_et_al_1999,Binney_and_Tremaine_2008}. The former was introduced in a series of papers in the 1980's \citep[e.g.][]{White_1983,Tremaine_et_al_1984, Weinberg_1986, Weinberg_1989}. Specifically, \citeauthor{Weinberg_1986} (\citeyear{Weinberg_1986}; \citeyear{Weinberg_1989}), described in detail how the torque that drives a satellite's decay, which is primarily caused by a resonant interaction with a dipolar mode induced in the halo, and additionally by other higher-order modes, could reproduce extremely well the results of $N$-body simulations.

The global halo response theory has the attractive feature that it removes the need to arbitrarily set a cut-off scale of the drag by choosing a value of the Coulomb logarithm in Chandrasekhar's formula, which is derived for an infinite homogeneous medium. Other approaches borne out of statistical mechanics, such as the theory of linear response of gravitating systems \citep[][]{Bekenstein_1989,  Colpi_et_al_1998, Colpi_et_al_1999}, also have the same shortcoming of having a built-in divergence of the drag to cope with, requiring a cut-off scale to be imposed. On the other hand, all of the approaches are challenged by how to self-consistently account for the ``self-gravity of the response'', namely the fact that, be it a local wake or a global mode, the overdensity associated with it might have non-negligible self-gravity, which could affect the dynamical interaction with the perturber \citep{Weinberg_1986, Colpi_et_al_1999}. 
Past studies of the global halo response were borne out of simulations carried out with a very specific numerical technique for gravitational problems, the self-consistent field method \citep[SCF;][]{ Hernquist_et_al_1992, Hernquist_et_al_1995}, and were never reproduced with other techniques such as the tree-code. With the low resolution available more than 20 years ago in satellite-primary $N$-body studies (i.e. less than $10^5$ particles per object), it was never clear if global modes were missed by most $N$-body methods because of the high discreteness noise level. The contention between the SCF method results and the other methods extended also to the problem of bar-halo interaction, in which breathing halo modes excited at resonances, and back-reacting on the bar, were claimed to play a major role in the interaction, causing the bar to carve a core in a Navarro--Frenk--White \citep[NFW;][]{Navarro_et_al_1996} halo \citep[][]{Weinberg_et_al_2002}. These results have never been confirmed with tree-based gravity solvers, which have since become the standard technique employed in numerical cosmology over the past two decades, and have generated most of our understanding of structure formation and evolution in the CDM cosmogony \citep{Kuhlen_et_al_2012}.  Published analysis of induced halo modes with SCF methods suggested that, for traditional particle-based gravitational solvers such as tree-codes, at least several hundred million particles would be needed in the primary halo to capture the resonances responsible for most of the torquing and energy transfer.  However, these conclusions were disputed by later $N$-body work performing convergence studies at increasing resolution \citep[][]{Dubinski_et_al_2009}.

With the many-fold increase in resolution that is possible with modern tree-codes running on the fastest supercomputing facilities, the time is ripe to tackle this challenging problem once again. We remark that the purpose of doing this is not only because one wants to understand the theoretical grounds of the orbital decay process in a robust way, given its key role in galaxy formation, but also because one wants to make sure that large-scale cosmological simulations do capture sufficiently well the orbital dynamics of satellites and produce trustworthy results.

In this work, we revisit in depth the  physical nature of the satellite-primary halo interaction, attempting to determine whether or not global halo modes are triggered as a result of the perturbation of the satellite. The second objective that we focus on is to find out if such global modes, when present, contribute significantly or not to drain the orbital energy and orbital angular momentum of the satellite, and hence if they are important to understand the nature of the sinking satellite problem. In doing so, we attempt to study the torques exerted on the satellite, separating the contribution of a local wake, which can be traced back to the more conventional interpretation of dynamical friction, from that of global modes. For realistic satellites, tidal heating and tidal mass loss represent another complication, as they affect the orbital decay by reducing the perturbing mass, so that the decay rate is significantly reduced \citep[][]{Colpi_et_al_1999, Taylor_et_al_2001, Taffoni_et_al_2002}. The opposite phenomenon can also occur, wherein gas inflows can increase the baryonic mass of the perturber, further complicating the picture \citep[e.g.][]{VanWassenhove_et_al_2014,Capelo_et_al_2015}. In order to understand more cleanly the nature of the torques, we will employ both live satellites, which can undergo tidal heating and mass loss, and rigid satellites. Note that all past work in the context of the global halo modes scenarios has been carried out with rigid satellites \citep[e.g.][]{Weinberg_1986,Weinberg_1989}. 

For this purpose, we simulate a merger between a Milky Way (MW)-like galaxy and a dwarf galaxy satellite, for which only the DM halo is modeled, and whose mass is 10 times smaller than that of the primary. Note that such a satellite halo has a mass compatible with the expected halo mass of the Large Magellanic Cloud (LMC) prior to infall, based on abundance matching constraints for a galaxy with the current stellar mass of the LMC \citep[e.g.][]{Li_et_al_2009}. Furthermore, the Gaia-Enceladus stream in phase space and other similar structures discovered by Gaia \citep[e.g.][]{Helmi_et_al_2018} suggest that a merger between the MW and a similarly massive satellite might have occurred in the past. Finally, we also investigate the impact of numerical resolution on the DM modes. In the past years, a handful of studies have investigated DM modes \citep[][]{Choi_et_al_2009, Ogiya_et_al_2016, Garavito_et_al_2019, Cunningham_et_al_2020}, but at a significantly lower mass resolution than what we achieve here. Also, the studies by \citet{Garavito_et_al_2019} and \citet{Cunningham_et_al_2020} performed a detailed analysis of the LMC and its orbit within the MW halo, and argued that a manifestation of the global modes could be detected in the stellar halo by Gaia, yet they did not carry out a theoretical study of the halo response, nor did they clearly identify  the global and local response of the halo or studied their role in the orbital decay of a satellite.

In order to avoid confusion in the terminology, we would like to clarify from the beginning that we will refer to the local response  of the halo as the {\it local wake}, and to the global response as {\it global modes}. More specifically, the local wake will be identified with the overdensity trailing the satellite, with a more quantitative definition being provided later on as we present our analysis, whereas global modes will correspond to density perturbations at scales larger than that of the wake, and will be quantitatively identified via the analysis of the halo density power spectrum. Note that such global modes will appear as a global distortion of the halo mass distribution excited by the satellite and, as such, are not simply trailing the satellite's orbital path.


\section{Numerical Setup}\label{sec:Numerical_setup}

This section describes the main-halo and satellite models, as well as the orbits of the simulations.

\begin{deluxetable*}{cccccc}
\tablecaption{Particle specifications for the live satellite simulations \label{tab:runs}} 
\tablehead{\colhead{Resolution:}& HR & MR & LR & ULR & }
\startdata
$N_{\rm DM}$,  $N_{\rm DM,sat}$ & $2.00 \times 10^8$, $2.00 \times 10^7$ & $2.50 \times 10^7$, $2.50 \times 10^6$ & $3.13 \times 10^6$, $3.13 \times 10^5$ & $3.91\times 10^5$, $3.91\times 10^4$ & \\
$N_{\star}$, $N_{\star \rm{, B}}$ & $5.00 \times 10^7$, $1.36 \times 10^7$ & $6.25 \times 10^6$, $1.70 \times 10^6$ & $7.81 \times 10^5$, $2.13 \times 10^5$ & $9.77 \times10^4$, $2.66\times 10^4$ & \\
$m_{\rm DM}$, $m_{\rm DM,sat}$ & $5.33 \times 10^3$, $5.52 \times 10^3$ & $4.26 \times 10^4$, $4.42 \times 10^4$ & $3.41 \times 10^5$, $3.53 \times 10^5$ & $2.73 \times 10^6$, $2.83\times 10^6$  & [M$_\odot$]\\
$m_{\star}$, $m_{\star {\rm , B}}$ & $715$, $714$ & $5.72 \times 10^3$, $5.71 \times 10^3$ & $4.58 \times 10^4$, $4.57 \times 10^4$ & $3.66 \times 10^5$, $3.66 \times 10^5$  & [M$_\odot$]\\
$\epsilon_{\rm DM}$, $\epsilon_{\rm DM,sat}$& $25$, $25$ & $50$, $50$ & $100$, $100$ & $200$, $200$  & [pc]\\
$\epsilon_{\star}$, $\epsilon_{\star}$& $15$, $15$ & $30$, $30$ & $60$, $60$ & $120$, $120$  & [pc]\\
\enddata
\tablecomments{Particles' numbers (denoted by $N$, with the subscripts $N_{\rm DM}$ for DM particles of the main halo, $N_{\rm DM, sat}$ for DM particles of the satellite, $N_{\star}$ for stellar disk particles, and $N_{\star,{\rm B}}$ for stellar bulge particles), masses (denoted by $m$, with the same subscripts as before), and softenings (denoted by $\epsilon$, with the same subscripts as before), for the four resolution levels: HR, MR, LR, and ULR. At each increasing resolution level, the particle mass (softening) is one eighth (one half) of that of the previous level. In the first row, the first number refers to the main halo and the second to the satellite. In the second row, the first number refers to the main-halo stellar disk and the second to the main-halo bulge.}
\end{deluxetable*}

\begin{deluxetable*}{cc|cc|cc||cc|cc}
\tablecaption{Galaxy models} \tablecolumns{4}
\tablehead{ \multicolumn{6}{c||}{Milky Way (MW)}  &  \multicolumn{4}{c}{Satellite} \\DM halo&   & Disk &  & Bulge & & Halo & & Rigid &}
\startdata
$M_{\rm vir}$ & $1.00 \times 10^{12}$  & $M_{\rm d}$ & $3.58 \times 10^{10}$ & $M_{\rm b}$ & $9.72 \times 10^{9}$ & $M_{\rm vir,sat}$ & $0.10 \times 10^{12}$ & $M_{\rm R}$ & $0.10 \times 10^{12}$ \\
$R_{\rm vir}$ & $258.00$ & $R_{\rm d}$ & $2.84$ & $n$ & $1.28$ & $R_{\rm vir,sat}$ & 71.12  & $\epsilon_{\rm R, L}$ & 3.40  \\
$c_{\rm vir}$ & 12.00 & $z_{\rm d}$ & $0.43$ & $\rho_{\rm b}$ & $ 1.17 \times 10^{10}$ &  $c_{\rm vir,sat}$ & 7.64  & $\epsilon_{\rm R, S}$ & 1.10 \\
&   & $\sigma_{0}$ & $129.10$  & $R_{\rm e}$ & $ 0.56$ &  &  \\
\enddata
\tablecomments{Parameters of the DM halos of the MW and the satellite, as well as of the stellar component of the MW, given in units of M$_\odot$, kpc, M$_\odot$~kpc$^{-3}$, and km~s$^{-1}$.}
\label{tab:ICfiles}
\end{deluxetable*}


\subsection{Galactic models}\label{sec:galactic_models}

In this work, we use a high-resolution model of the MW-satellite system (with $\sim$0.3 billion particles), as well as lower-resolution copies for convergence studies. A resolution summary of the different models applied in this work can be seen in Table~\ref{tab:runs}. With the development of \textsc{Pkdgrav3} \citep[][]{Potter_et_al_2017}, we can now explore the $\mathcal{O}(10^8)$ regime of DM particles in isolated $N$-body simulations. Consequently, one aim of this paper is to show conclusive results about the needed resolution of DM models and their relation to dynamical friction. All our self-consistent $N$-body models were generated with \textsc{GalactIC's}, which adopts the methods described in \citet{Kuijken_Dubinski_1995}, \citet{Widrow_Dubinski_2005}, and \citet{Widrow_et_al_2008}. Our initial conditions reproduce a MW-like galaxy, which contains a stellar disk and bulge, and a DM halo, and a dwarf satellite which consists purely of DM. The particular values given in the following subsections, in which we give an overview of the individual components, are taken from \citet{Widrow_Dubinski_2005} in \textsc{GalactICs} units and converted to physical units to provide the reader a physical intuition.


\subsubsection{The Milky Way}

The DM halo density profile follows the NFW profile,

\begin{equation}
    \rho(r) = \frac{\rho_{\rm s}} {(r/r_{\rm s}) (1 + r/r_{\rm s})^{2}},
\end{equation}

\noindent  where $\rho_{\rm s}$ denotes four times the density at the scale radius $r_{\rm s}$. We assume an MW model at $z = 0$, with a virial mass $M_{\rm vir} = 1.00 \times 10^{12}$~M$_\odot$ and a concentration $c_{\rm vir} = 12$ \citep[similarly to what was done in][]{Villalobos_et_al_2008}. Furthermore, we assume that the NFW profile is exponentially truncated at $r_{\rm cutoff} = 310$~kpc.

We model the stellar disk with an exponential profile in the direction of the projected radius $R$ and with a sech$^2$ profile in the $z$-direction \citep[][]{Spitzer_1942,Binney_and_Tremaine_2008}:

\begin{equation}
    \rho_\star (R, z)=  \frac{M_d}{4 \pi R_d^2 z_d} \exp \left (-\frac{R}{R_{\rm d}} \right )  {\rm sech}^2 \left (\frac{z}{z_{\rm d}} \right)
\end{equation}

\noindent where $R_{\rm d}=2.84$~kpc denotes the disk scale radius in cylindrical coordinates, $M_{\rm d} = 3.58 \times 10^{10}$~M$_{\odot}$ the total disk mass, and $z_{\rm d} =0.43$~kpc the disk scale height \citep[similar to the MWb disk model in][]{Widrow_Dubinski_2005}. Additionally, we choose a radial velocity dispersion at the galaxy center of $\sigma_0 = 129.1$~km~s$^{-1}$. Furthermore, we also model a stellar bulge according to

\begin{equation}
\rho_{\rm b} (r) = \rho_{\rm b} \left(\frac{r}{R_{\rm e}}\right)^{-p} \exp \left( -\kappa \left(\frac{r}{R_{\rm e}} \right ) \right)^{1/n},
\end{equation}

\noindent with \citep[similarly to what was done in][]{Widrow_et_al_2008} $\rho_{\rm b} = 1.17 \times 10^{10}$~M$_\odot$~kpc$^{-3}$ being a normalization constant\footnote{Note: in the actual code, $\rho_{\rm b}$ is calculated via $\sigma_{\rm b} = \{4 \pi n b^{n(p-2)} \Gamma[n(2-p)]R_{\rm e}^2 \rho_{\rm b}\}^2$, with $\sigma_{\rm b}^2$ the depth of the gravitational potential and $\Gamma$ the Gamma function.}, $R_{\rm e} = 0.56$~kpc denoting the effective radius,  $n= 1.28$ the \citet{Sersic_1963,Sersic_1968} index, and $\kappa = 1.9992 \,n - 0.3271$ \citep[][]{Capaccioli_1989}. Setting $p = 1- 0.6097 / n  + 0.05563/n^2$ \citep[][]{Marquez_et_al_2000} leads to the S{\'e}rsic law for a projected density profile \citep[see][]{Prugniel_Simien_1997, Terzic_Graham_2005}. A summary of all values used for the MW model can be seen in the left-hand side of Table~\ref{tab:ICfiles}. For our high-resolution (HR) runs, the MW analogue is sampled by $2 \times 10^8$ DM particles and $5 \times 10^7$ stellar particles, thus setting the DM and stellar particle mass to $5.33 \times 10^3$ and 715~M$_{\odot}$, respectively. For our medium-resolution (MR), low-resolution (LR), and ultra-low-resolution (ULR) runs, we simply multiplied by 8, 64, and 512, respectively, the corresponding HR particle-mass values (see Table~\ref{tab:runs}).


\subsubsection{Satellite}

We model the satellite in two different ways. In the first case -- the ``live'' satellite -- we employ an NFW profile for the DM component, with $M_{\rm vir,sat} \simeq 0.1 $~M$_{\rm vir}$, $r_{\rm vir,sat} = 71.12$~kpc, and $c_{\rm vir,sat}=7.64$ \citep[see][]{Bullock_et_al_2001}. These values are consistent with a $z = 1$ system, as done in, e.g. \citet{Villalobos_et_al_2008}. For our HR runs, the live satellite is sampled by $2 \times 10^7$ DM particles, thus setting the DM particle mass to $5.52 \times 10^3$~M$_{\odot}$. The DM and stellar softenings \citep[Dehnen K1 kernel;][]{Dehnen_2001} were set (for both the main halo and the satellite) to 25 and 15~pc, respectively, following the scaling by \citet{Kazantzidis_et_al_2005}. For our MR, LR, and ULR runs, we simply multiplied by 2, 4, and 8, respectively, the corresponding HR softening values (see Table~\ref{tab:runs}).

In the second case -- the ``rigid'' satellite -- we employ a single stellar particle of mass $M_{\rm R}=1.10 \times 10^{11}$~M$_\odot$, with a softening of $\epsilon_{\rm R, L} = 3.4$~kpc, which is similar to the effective radius of the LMC \citep[see][]{Colpi_et_al_1999}. We also considered a smaller softening $\epsilon_{\rm R, S} = 1.1$~kpc. The values used for these models are listed in Table~\ref{tab:ICfiles}.


\begin{deluxetable}{cccc}
\tablecaption{Orbits} \tablecolumns{4}
\tablehead{\colhead{$i$}& $x$,  $y$,  $z$ & $v_x$, $v_y$, $v_z$ \\ & [kpc] & [km~s$^{-1}$] }
\startdata
$0 ^\circ$ & $122.18$,  $0.00$,  $0.00$ & $-107.11$, $80.33$, $0.00$  \\
$30 ^\circ$ & $105.81$,  $0.00$,  $61.10$ & $-92.76$, $80.33$, $-53.56$  \\
$60 ^\circ$ & $61.09$, $0.00$,  $105.81$ & $-53.57$, $80.33$, $-92.76$ \\
\hline
$90 ^\circ$ & $0.00$,  $0.00$,  $122.18$ & $0.00$, $0.00$, $-107.11$\\
\enddata
\tablecomments{Initial orbital parameters for the four orbits (Section~\ref{sec:orbits}); the first three rows depict similar orbits, which differ only in the inclination. For the $i = 0^\circ, 30^\circ$, and $60^\circ$ cases, the initial values are obtained via $x = d \cos(i)$, $y = 0$, $z = d \sin(i)$, $v_x = - v_{\rm r, sat}\cos(i)$, $v_y = v_{\theta, sat}$, and $v_z = - v_{\rm r, sat} \sin(i)$. For the $i = 90^\circ$ simulation, we model a purely radial encounter, with $v_x = v_y = 0$.}
\label{tab:orbits}
\vspace{-30pt}
\end{deluxetable}

\subsection{Orbits}\label{sec:orbits}

In this section, we present the orbits that we used for our merger simulations. We first make a standard choice for the orbits of the satellite in the primary MW halo, and then we also consider an orbit for which the decay curve is expected to be different. The reason for this is that we do not want to restrict ourselves to the most probable orbits given by cosmological simulations, but also show how the results that we will extract about the halo response to the satellite's perturbation are general rather than dependent on the orbital parameters.

We model our collisions with four different inclination angles: $i = 0^\circ, 30^\circ, 60^\circ$, and $90^\circ$, where $0^\circ$ means that the orbit starts parallel to the MW disk mid-plane. The individual satellite is placed on a prograde orbit at a distance $d = 122.2$~kpc from the center of the MW, which is the redshifted virial radius of the MW at $z = 1$ for a virial mass of $M_{\rm vir} = 0.57 \times 10^{11}$~M$_\odot$ \citep[this is a procedure similar to that in][]{Villalobos_et_al_2008}. For the initial velocity parameters, we choose the peak of the velocity distribution according to \cite{Benson_2005}, which corresponds to $v_{r} = 0.8\,v_{\rm vir,1}$ and $v_{\theta} = 0.6\,v_{\rm vir,1}$, with $v_{\rm vir,1}$ denoting the circular velocity at the virial radius at $z=1$. We assume that the interaction between the two galaxies began at $z=1$ and therefore we obtain $v_{\rm vir,1} = 133.89$~km~s$^{-1}$ \citep[see][]{Villalobos_et_al_2008}, thus obtaining $v_{r} = 107.11$ and $v_{t} = 80.33 $~km~s$^{-1}$. For the detailed values of the orbital parameters, we refer to Table~\ref{tab:orbits}.

The previously mentioned exception that we choose in this project is to neglect $v_{\theta}$ and model a purely radial encounter with an inclination angle of $i = 90^\circ$, which means that we place our satellite in an orbit perpendicular to the MW's disk. This leads to an eccentricity of $e = 1$ and thus it is not comparable with the previously described orbits.


\section{Results}\label{sec:Results}

In this section, we present our results, which include an analysis of live- and rigid-satellite simulations, for different orbital parameters and at different resolutions. In the following three sections, we describe in detail the DM global modes (Section~\ref{sec:DMglobal}), DM local wakes (Section~\ref{sec:local_wake}), and torques (Section~\ref{sec:Torques}) for three orbital configurations ($i = 0^\circ, 30^\circ$, and $60^\circ$) described in Section~\ref{sec:orbits}.

\begin{figure}[htb!]
\includegraphics[width=0.5\textwidth]{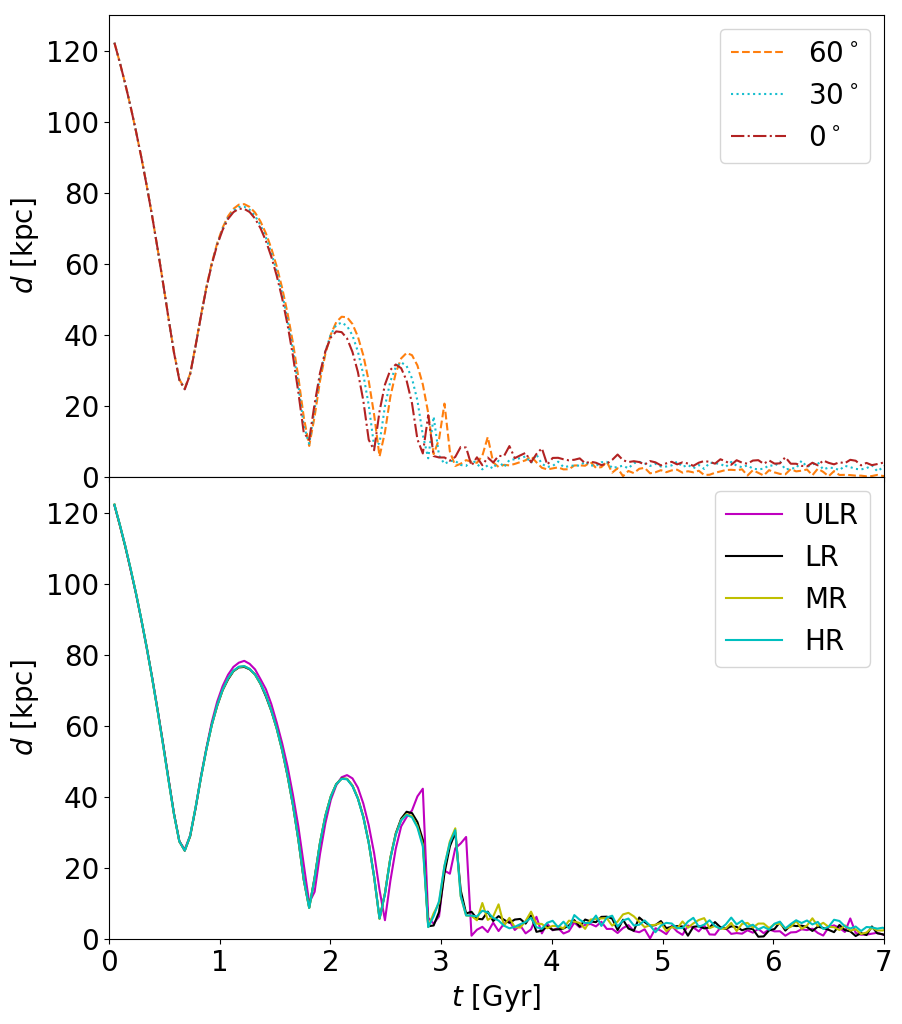}
\vspace{12pt}
\caption{
\textit{Top panel}: Distance as a function of time between the center of mass of the bound live-satellite particles and the center of mass of the MW halo, for three HR runs ($0^\circ$, $30^\circ$, and $60^\circ$). \textit{Bottom panel}: Same as the top panel, but comparing the four different resolution runs of the $60^\circ$ simulation. All curves (of both panels) depict similar, although slightly different, orbital histories.
}
\vspace{12pt}
\label{fig:distance}
\end{figure}

In Figure~\ref{fig:distance}, we show the orbital decay of the live satellite of these HR simulations (top panel), along with the comparison at different resolutions for the $i = 60^\circ$ run (bottom panel). The orbits are all very similar, owing to the spherical symmetry of the main DM halo and to the same radial and tangential velocities. The excellent resolution convergence is due to the fact that we always resolve the main physical mechanisms affecting the orbital decay, as explained below.

\subsection{Dark matter global modes}\label{sec:DMglobal}

\begin{figure*}[htb!]
\includegraphics[width=\textwidth]{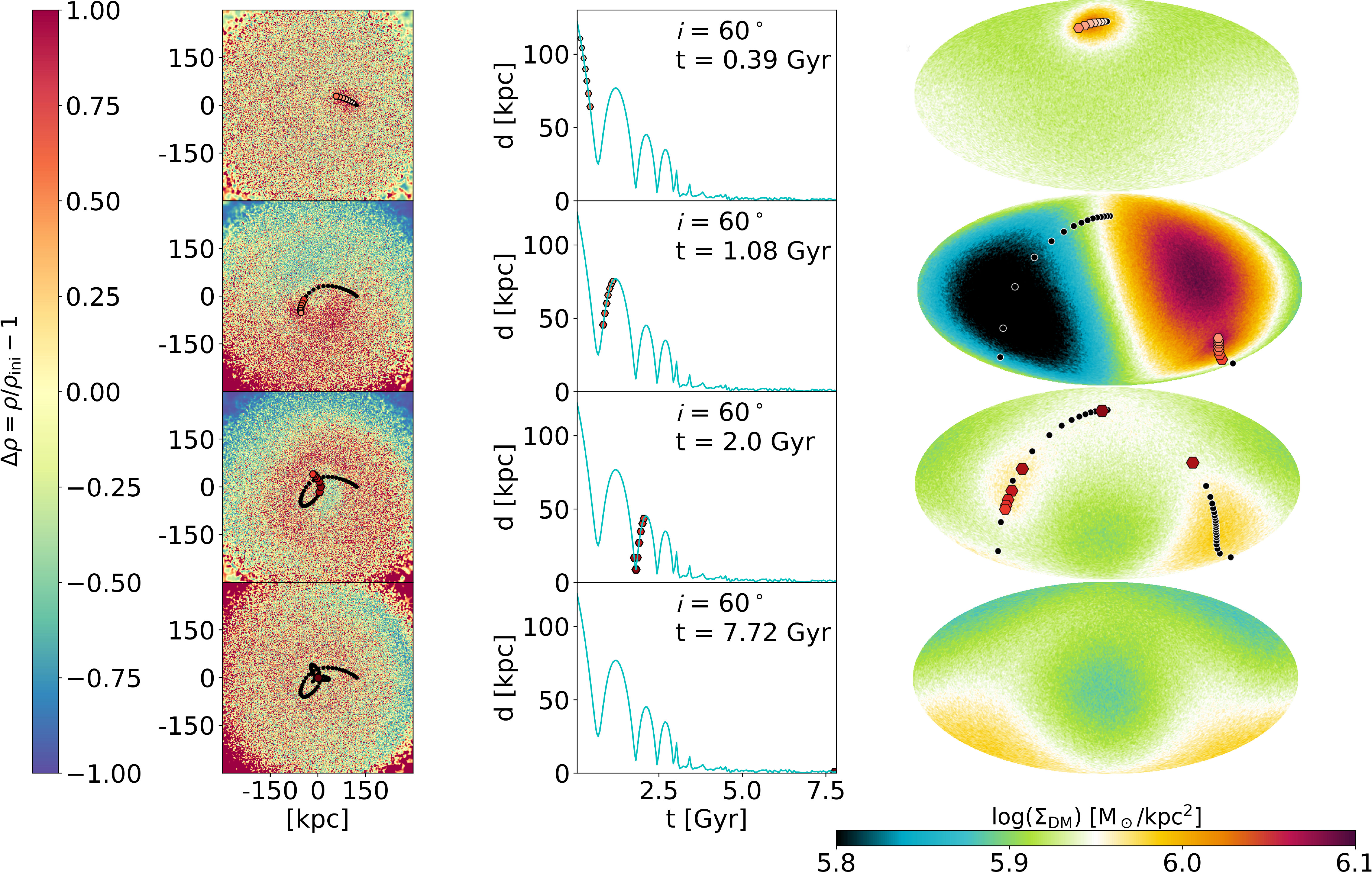}
\vspace{2pt}
\caption{The $i = 60^\circ$ HR live-satellite simulation: the snapshot was centered on the DM of the MW and afterwards the satellite position was calculated (see distance plot). Once the halo was centered, only DM particles from the MW are kept and projected onto a sphere of radius $310$~kpc. Hence, the information about the distance from the center is lost and we can calculate the surface density in a $d \theta \times d \phi$ region in the sky. The \textit{right-hand panels} show the Mollweide projections onto this sphere at the beginning of the simulation, at first apocenter ($\sim$1.08~Gyr), at second apocenter ($\sim$2.0~Gyr), and at the end of the simulation ($\sim$7.72~Gyr), from top to bottom. The hexagons in the plots show the current and past six positions of the satellite with our time resolution of $\Delta t = 0.048$~Gyr. The bigger and redder the dots, the shorter the distance between the satellite and the center of the MW. The \textit{middle panels} depict the orbital decay curve and the same hexagons as in the Mollweide projections. In the \textit{left-hand panels}, we show $\Delta \rho_{\rm DM}:= \rho/\rho_{\rm ini}-1$ in the orbital plane (with a thickness of $10$~kpc for each slice), similar to what has been done in \citet{Garavito_et_al_2019}.}
\vspace{12pt}
\label{fig:mollweide60}
\end{figure*}

\begin{figure*}[htb!]
\includegraphics[width=1.0\textwidth]{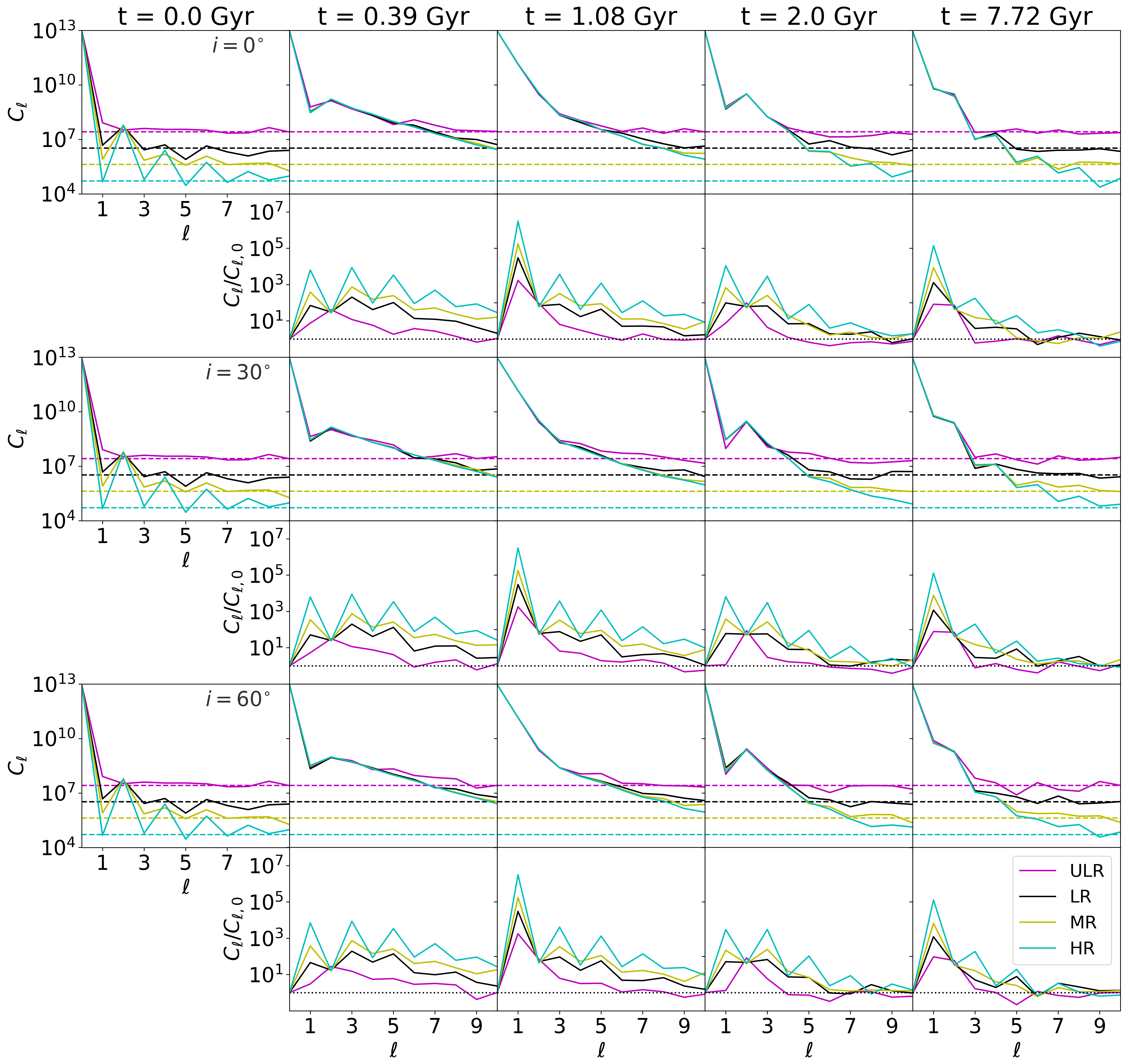}
\vspace{2pt}
\caption{This figure consist of six rows and five columns and shows the power spectrum analysis of our simulations. The figure is grouped such that always two rows belong to the same simulation; the $0^\circ$ simulations in rows 1 and 2, the $30^\circ$ simulations in rows 3 and 4, and the $60^\circ$ simulations in the last rows of our live-satellite runs with various resolutions (HR, MR, LR, and ULR). The \textit{first row} of each simulation always shows the (not normalized) power spectrum of the DM MW halo, with the first column showing the initial power spectrum. As predicted, nearly all power is in the $\ell = 0$ mode at time $t=0$ Gyr. The horizontal dashed lines show the theoretical sampling noise of the DM halo. The \textit{following columns} show the evolution of the power spectrum at different times of the merger (from left to right: $t=0.39$, $1.08$, $2.0$, and $7.72$~Gyr). All encounters excite the $\ell = 1$ in all resolutions, but higher modes are diminished if the resolution becomes too low. Especially, in the ULR simulation already the $\ell = 3$ and $\ell = 5$ mode cannot be seen properly. In contrast to this, in the HR simulations we can even see modes up to $\ell \approx 9$. The second row of each simulation block depicts the normalized power spectrum of the HR, MR, LR, and ULR runs. In that case we can see that the $\ell = 1$ mode of all HR runs is exited by $10^7$ times in contrast to the initial value at the beginning of the simulation. Furthermore, since we constructed our initial models with a self-consistent code, we can clearly see that, at all resolutions, the $\ell=2$ mode has the same value. This can be attributed to the fact the we removed the stellar disk for this analysis.}
\vspace{12pt}
\label{fig:powerspectrum_norm}
\end{figure*}

\begin{figure*}[htb!]
\includegraphics[width=1.0\textwidth]{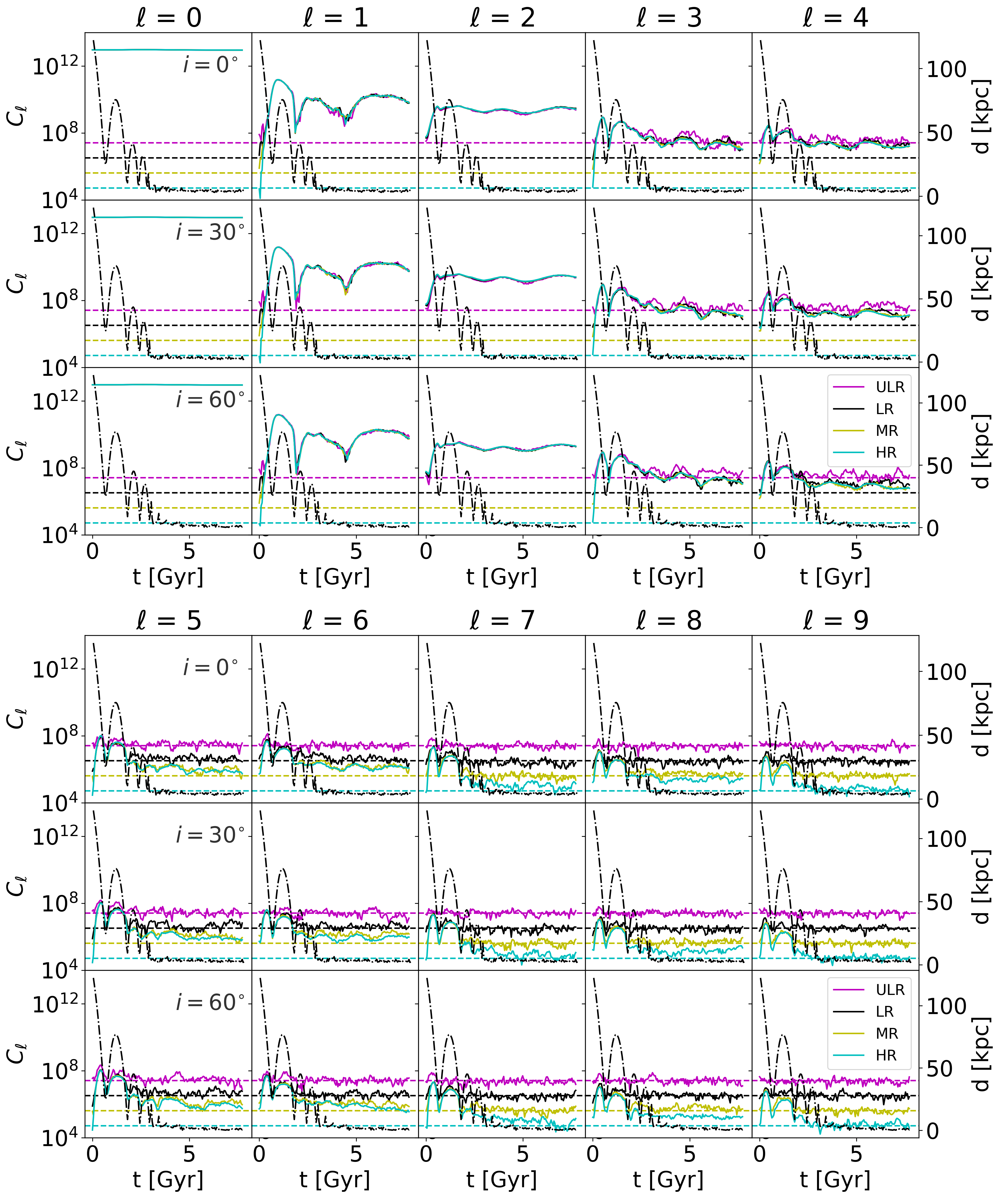}
\vspace{2pt}
\caption{\textit{Left-hand y-axis:} time evolution of each $\ell$ mode (\textit{increasing from left to right}) for the $0^\circ$ (rows 1 and 4), $30^\circ$ (2 and 5), and $60^\circ$ (3 and 6) simulations. \textit{Right-hand y-axis:} distance of the center of mass of the bound DM particles of the satellite to the center of the MW. It is clear that modes higher than $\ell = 5$ are only seen with increasing resolution. Indeed, the $\ell=9$ mode can be only clearly seen in the HR runs. Moreover, we can see that the impact angle $i$ does not affect the individual modes. In case of the $\ell=1$ mode we can see that it peaks at roughly $1$~Gyr, which is roughly half the crossing time at the MWs virial radius. }
\vspace{12pt}
\label{fig:powerspectrum_norm2}
\end{figure*}

In this section, we explore the response of the MW's DM halo to the perturbation caused by the live satellite. For this purpose, we produced \citet{Mollweide_1805} projections of the MW's DM halo's surface density\footnote{Using the python package \textsc{HEALPY} \citep{healpy1,healpy2}.}, shown in the right-hand panels of Figure~\ref{fig:mollweide60} for different times of the $60^{\circ}$ HR live-satellite case. For each projection, the DM halo was re-centered to the center of the MW DM, with the shrinking-sphere method \citep[][]{Power_et_al_2003}, and projected onto a sphere of $310$~kpc radius, which corresponds to the cutoff radius of the MW. With this setup, we visually found that the dipole moment is maximally excited at roughly the first apocenter (see the second column of Figure~\ref{fig:mollweide60}).

Our findings were confirmed by our power spectrum analysis. In contrast to the widely used overdensity power spectrum \citep[e.g. the power spectrum of the CMB; see ][]{Hinshaw_et_al_2013}, we calculated the spherical harmonic coefficients $c_{l,m}$ of the 310-kpc-sphere's surface density and summed over all $m$ values for a given $l$: $C_\ell := \sum_m |c_{l,m}|^2$. In this work, we refer to $C_\ell$ as a ``mode'' and simply denote it as $\ell$. In Figure~\ref{fig:powerspectrum_norm}, we show the power spectrum for the first $\ell = 9$ values, including various time steps as well as different orbits [$i=0^\circ , 30^\circ, \text{and}\; 60^\circ $]. We also include plots of a normalized power spectrum, which is obtained by dividing the power spectrum by the initial values obtained at the beginning of the simulation. Furthermore, in Figure~\ref{fig:powerspectrum_norm2}, we also show the time evolution of these modes and compare them against the orbital decay of the satellite. From these two Figures, we can draw the following conclusions:

\begin{itemize}

\item \textit{The global mode can be found at all resolutions:}\\
We observe that the $\ell = 1$ mode is excited by more than six orders of magnitude, in contrast to the initial state, at $t = 1.08$~Gyr, and more than five orders of magnitude at $t = 7.77$~Gyr in the HR simulation. Moreover, at those same times, we observe a similar $\ell = 1$ excitation at all resolutions.

\item \textit{The relative amplitude of the lowest-order global mode is mostly resolution-independent:}\\
The differences in the peaks' amplitudes of the $\ell=1$ mode
can be explained by the initial sampling noise at $t=0$ (see the left-hand panels in Figure~\ref{fig:powerspectrum_norm}). That is, we have checked that the amplitudes can be matched with one another if we re-normalize by the number of particles ($N$), which assumes a re-normalization of the surface density of $\sqrt{N}$ and simply comes from the Poissonian sampling noise. Since our models are axisymmetric,\footnote{Even though the DM halo is spherically symmetric, when we exclude the axisymmetric disk for our analysis, we break the spherical symmetry of the gravitational potential, thus exciting an ``artificial'' $\ell = 2$ mode (see Appendix~\ref{sec:RELAX}).} all $\ell$ modes larger than zero exist only due to the discrete sampling noise.

\end{itemize}

Figure~\ref{fig:powerspectrum_norm} also shows that the power spectrum for modes of order $\ell > 2$ is resolution-dependent, in the sense that higher-order modes are either barely at the noise level or below the noise level for the ULR and LR runs. Modes $\ell=3$ and higher become clearly discernible only at medium and high resolution. One should be reminded that the primary halo in the MR runs, with more than 20 million particles, is actually better resolved than galaxy-sized halos in any large cosmological simulation, being matched only by the highest-resolution cosmological zoom-in simulations of Milky Way-sized galaxies \citep[][]{Wetzel_et_al_2016, Sokolowska_et_al_2017, Hopkins_et_al_2018}. We will return to this point in Section~\ref{sec:Discussion}.


\subsection{Dark matter local wake}\label{sec:local_wake}

\begin{figure*}[htb!]
\includegraphics[width=\textwidth]{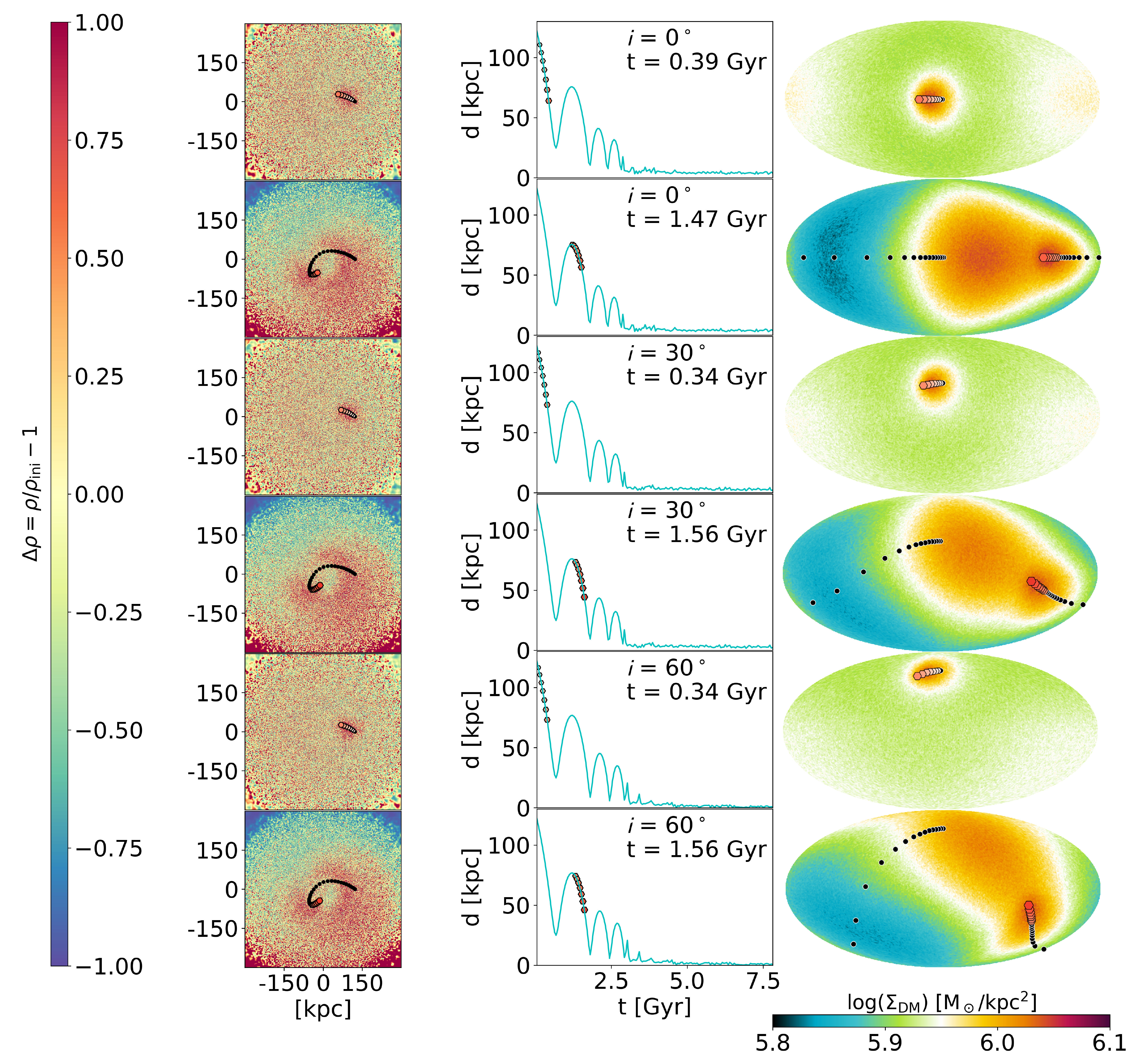}
\vspace{2pt}
\caption{The maximum of the local wake in the Mollweide projections of three of our live-satellite HR runs. From visual inspection, the maximum occurs always shortly before the first and second peri-center passage. From left to right: the overdensity $\Delta \rho$ for a 10 kpc thick slice around the satellites orbital plane, the distance from the center of mass of the bound satellite particles to the MW center, and the Mollweide projection of the DM halo of the MW. The hexagons in the plots depict the current and past six positions of the satellite with our time resolution of $\Delta t = 0.048$~Gyr. From top to bottom: $i = 0^\circ$ at $t = 0.39$ and $1.47$~Gyr, $i= 30^\circ$ and $i = 60^\circ$ at $t = 0.34$ and $1.56$~Gyr.}
\vspace{12pt}
\label{fig:localwake}
\end{figure*}

In the previous section, we showed that the $\ell = 1$ mode is excited at roughly the same level, at all resolutions. In this section, we attempt to find and identify the local wake, which corresponds to the classical description of dynamical friction \citep[][]{Chandrasekhar_1943, Tremaine_et_al_1984, Binney_and_Tremaine_2008}. The local wake, if present, should contribute to the torque responsible to drive the satellite's orbital decay. We know that the orbital decay curve is practically the same for all the inclinations and resolutions we analysed (see Figure~\ref{fig:distance}). Therefore, we can expect the local wake either to be prominent or negligible in all runs, so that it would contribute to a significant part of the torque or to a negligible part of it equally in all runs, or else one would expect to see a clear dependence on resolution of the orbital decay curve. In Figure~\ref{fig:localwake}, the visual maximum of the local wake is shown on the Mollweide projection. It is remarkable that the local wake is indeed visible, in spite of the fact that the projection of the DM halo density onto a large sphere is not ideal to depict it, since we lose the local 3D density of the local wake information when projecting.

\begin{figure*}[htb!]
\includegraphics[width=\textwidth]{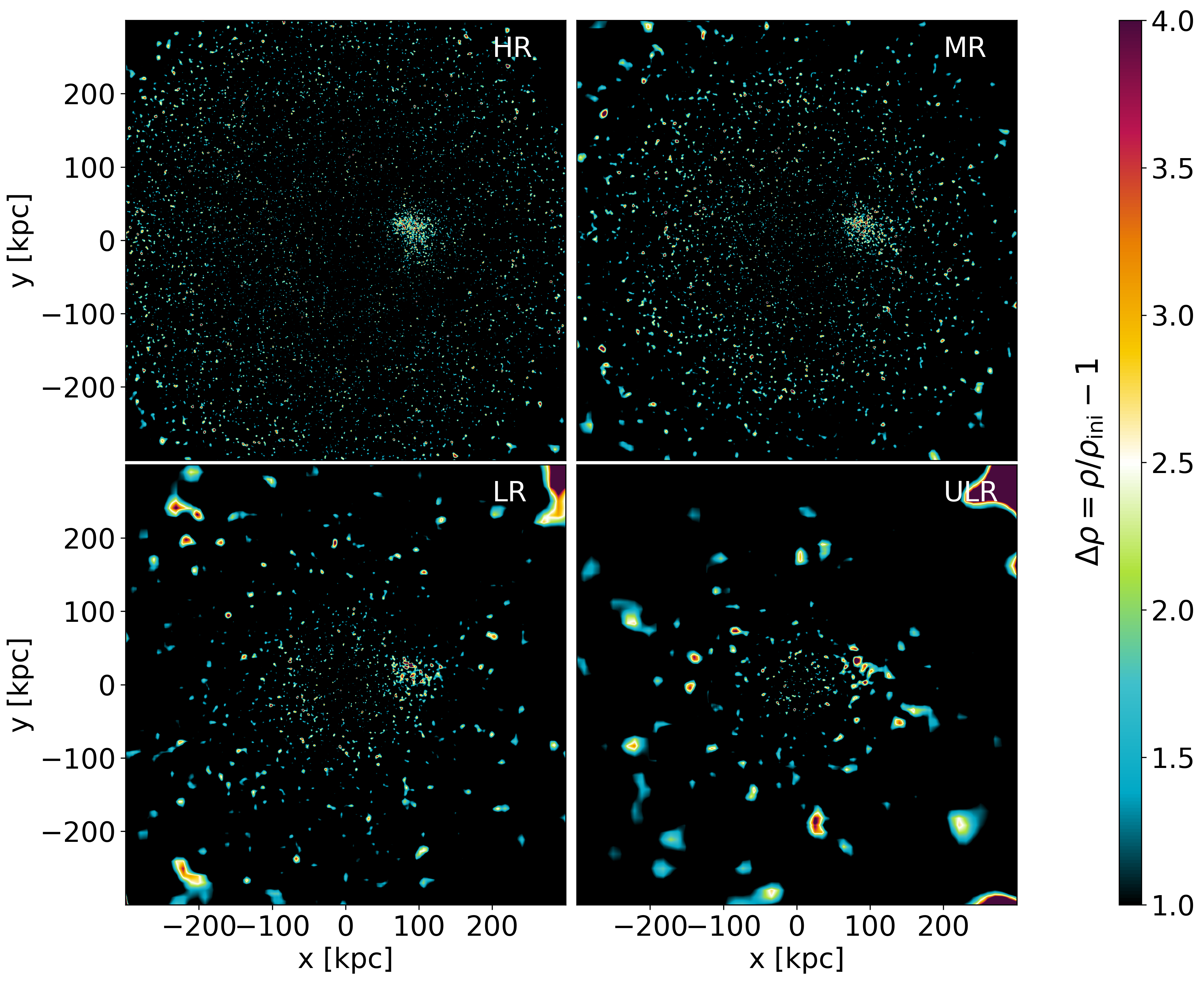}
\vspace{12pt}
\caption{The local wake in the satellite plane at $t = 0.34$~Gyr, for the four simulations of the live-satellite $60^{\circ}$ case (from top-left to bottom-right: HR, MR, LR, and ULR). The colorbar range starts at $\Delta \rho \geq 1$ and therefore the images depict all the regions that at least doubled their density value in contrast to the initial NFW distribution. It is evident that, as the resolution decreases, the local wake extent and the extent of the numerical noise are becoming comparable. Therefore, it is increasingly difficult to constrain the size of the local wake.}
\vspace{12pt}
\label{fig:binney_tremaine}
\end{figure*}

In Figure~\ref{fig:binney_tremaine}, we followed the analysis of \citet{Garavito_et_al_2019}, albeit with small modifications: we cut a cylinder of roughly $10$~kpc height around the orbital plane of the satellite, depict the 3D overdensity $\Delta \rho_{\rm DM}  = \rho_{\rm DM} / \rho_{\rm DM_{ini}} -1 $, with $ \rho_{\rm DM_{ini}}$ denoting the initial volume density at the beginning of the simulation, and compare our four resolutions of the $i=60^\circ$ live-satellite simulation. We are aware of the fact that the choice of $\Delta \rho_{\rm DM}$ leads to a non symmetric value range -- $[-1 - \infty)$ -- but our choice is historically motivated (see Section \ref{sec:Torques}). For simplicity, we chose, as the plane for the slice, the initial orbital parameters of the satellite. However, we have investigated the simulated orbit of the satellite and found that the orbit deviates by only a as small amount from this plane until the satellite is disrupted. Inspecting the four panels of Figure~\ref{fig:binney_tremaine}, it is easy to visually depict the local wake. However, when reducing the resolution, it becomes increasingly difficult to distinguish the local wake from the simulation noise. Indeed, in the ULR simulation, the local wake can be only identified with the help of the other simulations. As done in the previous section, we also investigate the power spectrum (Figure~\ref{fig:powerspectrum_norm}). The radial extension of the local wake is difficult to estimate from the power spectrum, since we might observe combinations of local and global modes. Furthermore, the size of the local wake changes as time passes by and finally, once the satellite settles at the center, it diminishes. However, we know from the power spectrum that the modes $\ell = 1, 3$, and $5$ are persisting, in the HR runs, until the very end of the simulation, which is much longer than the sinking time of the satellite. Therefore, any local contribution that is not in superposition with the global modes must be at least an $\ell=7$ mode. In fact we can, except in the ULR simulation, observe an excited $\ell = 7$ mode in all simulations at the first peri-center.

\begin{figure}[htb!]
\includegraphics[width=0.4\textwidth]{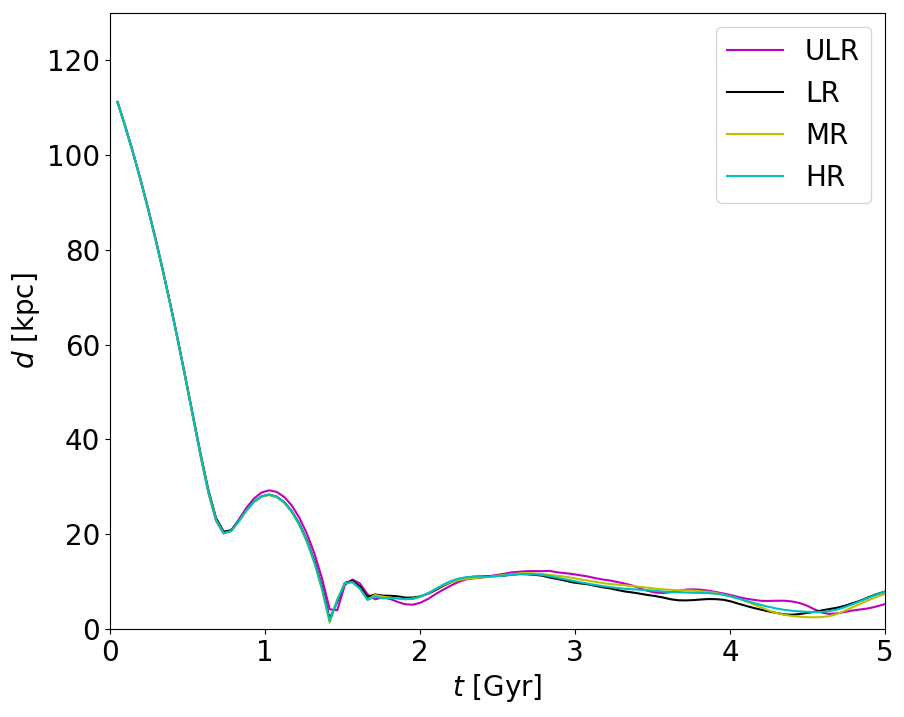}
\vspace{12pt}
\caption{Distance as a function of time between the rigid-satellite particle and the COM of the MW halo, for the four different resolution runs of the $60^\circ$ simulation. All curves depict very similar orbital histories.}
\vspace{12pt}
\label{fig:distance_resolution_point}
\end{figure}

\subsection{Torques: global mode versus local wake}\label{sec:Torques}

In the previous two sections, we showed that we have a local wake as well as a global mode in all our simulations, both triggered by the satellite. Furthermore, to our surprise, we can observe the global mode at every resolution. Therefore, it is important to investigate the orbital decay pattern for each simulation and explain where the main torque comes from. For this purpose, we ran the rigid-satellite simulations, using a point mass as the satellite, which allows us to accurately compute the torque and angular momentum of the satellite and ignore additional effects such as mass loss or deformation \citep[see][]{Taylor_et_al_2001}. In Figure~\ref{fig:distance_resolution_point}, we show the orbital decay of the rigid satellite with respect to the center of mass of the entire snapshot\footnote{We note that we did not use the shrinking-sphere method nor the potential-minimum method, since both methods did not yield the correct torques.} (hereafter COM). It is again indisputable that all simulations depict the same orbital decay, although they significantly differ from the orbits of the live satellites, in good agreement with previous work \citep[e.g.][]{Colpi_et_al_1999, Taylor_et_al_2001}, which explains this with the lack of mass loss and the compactness of the satellite.

In the following section, we analyze the exerted torque on the satellite rather than the force itself, since this allows us a direct comparison with the orbital decay. We define the projected torque from the $i$-th particle on the satellite via

\begin{equation}
\tau_{{\rm proj,} i} := \vec{\tau}_i \cdot \frac{\vec{J}_{\rm sat}}{|\vec{J}_{\rm sat}|},
\end{equation}

\noindent where $\vec{J}_{\rm sat} =  M_{\rm sat}(\vec{r}_{\rm sat} \times \vec{v}_{\rm sat} )$, $M_{\rm sat}$, $\vec{r}_{\rm sat}$, and $\vec{v}_{\rm sat}$ are the satellite's angular momentum, mass, position, and velocity, respectively (all computed with respect to the COM), and

\begin{equation}
\vec{\tau_i} =   \vec{r}_{\rm sat} \times  \frac{G m_i M_{\rm sat}} { (\vec{r}_{\rm sat} - \vec{r}_{i})^3} (\vec{r}_{\rm sat} - \vec{r}_{i}),
\end{equation}

\noindent where $G$ is the gravitational constant, denotes the torque of the $i$-th particle, of mass $m_i$ and position $\vec{r}_i$, which is at least one softening length ($\epsilon_{\rm sat}$) away from the satellite. We then compute the total projected torque on the satellite as $\tau_{\rm proj, tot} = \sum_i {\tau}_{{\rm proj,} i} = \sum_i \vec{\tau}_i \cdot {\vec{J}_{\rm sat}}/{|\vec{J}_{\rm sat}|}$.

\begin{figure}[htb!]
\centering
\includegraphics[width=0.5\textwidth]{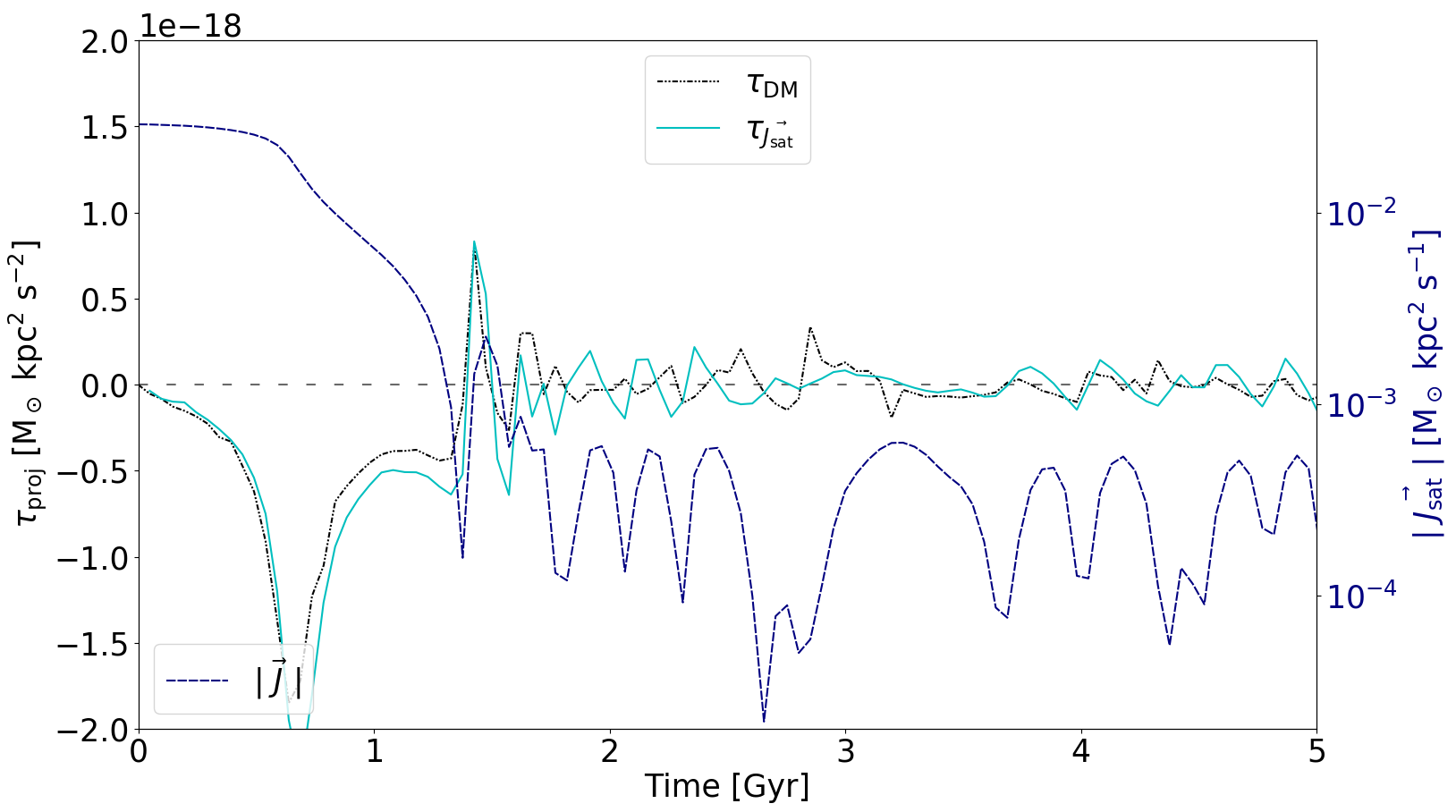}
\vspace{12pt}
\caption{The left-hand $y$-axis depicts the instantaneous projected torque on the rigid satellite coming from DM (black, dash-dotted line) and the time-averaged projected torque from the actual angular momentum of the satellite (turquoise, solid line). The horizontal, loosely dashed, black line shows the border between positive and negative torques, which is calculated with respect to the instantaneous angular momentum vector of the satellite. For this purpose, we centered the entire simulation on the COM of the entire snapshot. The right-hand $y$-axis shows the total angular momentum of the satellite (blue, dashed line).}
\vspace{12pt}
\label{fig:tot_torque}
\end{figure}

\begin{figure*}[htb!]
\centering
\includegraphics[width=0.8\textwidth]{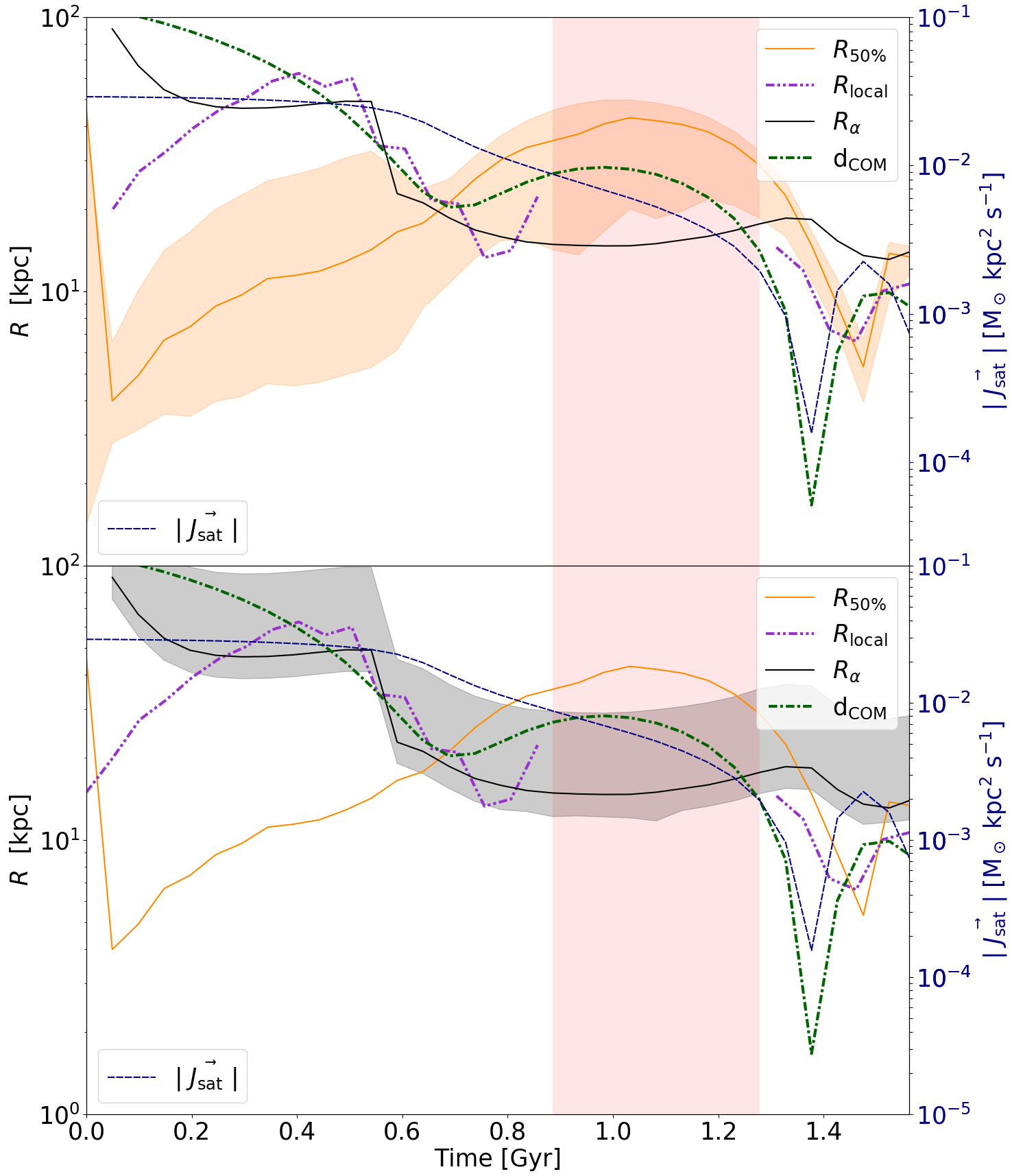}
\vspace{12pt}
\caption{The left-hand $y$-axis shows a radius or distance in kpc as a function of time. The solid, orange line depicts $R_{50 \%}$, the double-dot-dashed dark violet line shows $R_{\rm local}$ (as described in Section~\ref{sec:Results}), and the dot-dashed dark green line is the distance from the COM. Furthermore, on the right-hand $y$-axis, we show the corresponding angular momentum of the rigid satellite (blue dashed line). Finally, the black solid line (labeled as $R_\alpha$) depicts the calculated local wake radii from \citet{Mulder_et_al_1983} with $\alpha = 8$ until $\approx 0.6$~Gyr and $5$ afterwards (see Equation~\ref{eq:overdensity}). The \textit{top panel} additionally shows an orange shaded region, which depicts $R_{25 \%}$ at the lower boundary and $R_{75 \%}$ at the upper boundary. The \textit{bottom panel} portrays a gray shaded region, which indicates the impact of our contour-threshold in the smoothed overdensity plots (upper and lower boundary for $\Delta \rho = 0.5$ and $\Delta \rho = 1.2$, respectively). Lastly, the vertical pink region outlines the time in which we cannot disentangle the global mode from the local wake since they are in superposition. For this reason, $R_{\rm local}$ is also interrupted in this part. It is evident that at the beginning the local wake estimate and the actual extent of the wake in the simulation ($R_{\rm local}$ vs. $R_\alpha$) significantly deviate. The reason for this behaviour is that we start our simulation well within the MW's halo, whereas in a real galactic setup a satellite would have not entered the main halo without already accumulating a local wake and therein at the beginning we underestimate the local wake radius. On the other hand, at the end of the simulation the theoretical prediction clearly overestimates the local wake extent.}
\vspace{12pt}
\label{fig:halfcumulative-time-series}
\end{figure*}

Figure~\ref{fig:tot_torque} shows the total projected torque on the satellite from DM particles, together with the satellite's angular momentum. Furthermore, since these values are calculated from instantaneous quantities, we also include the time-averaged torque from the angular momentum variation,

\begin{equation}
\tau_{\vec{J}_{\rm sat}}(t) : = \frac{|\vec{J}_{\rm sat}(t)| - |\vec{J}_{\rm sat}(t-\Delta t)|}{\Delta t},
\end{equation}

\noindent where $\Delta t = 0.048$~Gyr is the time between two consecutive snapshots. Using these definitions, we can now also calculate the cumulative projected torque as a function of distance from the satellite: $\tau_{\rm proj}(d) = \sum_{r = 0} ^{d}\tau_{{\rm proj,} i}(r)$, where $r$ is the distance from the satellite. Since we wish to know where the main contribution of the torque is coming from, we then calculate the radii $d$ at which the cumulative torque $\tau_{\rm proj}(d)$ reaches $25\%$, $50\%$, and $75\%$ of $\tau_{\rm proj,tot}$, which we call $R_{25\%}$, $R_{50\%}$, and $R_{75\%}$, respectively. We calculate $R_{50\%}$ for every time-step and plot it as the solid orange line in Figure~\ref{fig:halfcumulative-time-series} with $R_{25\%}$ and  $R_{75\%}$ confining the shaded orange region. In this figure, $R_{50\%}$ slowly rises and reaches a maximum value of roughly $\approx 40$~kpc. As time proceeds, the difference between $R_{25\%}$, $R_{50\%}$, and  $R_{75\%}$ becomes considerably smaller.

In order to estimate the local wake in a quantitative way, we smoothed the overdensity plot and used a threshold of $\Delta \rho_{\rm DM} \geq 0.8$ at the beginning and $\Delta \rho_{DM} \geq 1.0$ at later times, to obtain the contour of the local wake (see Figure~\ref{fig:binney_tremaine_torques_point}). These values have been chosen such that the local wake can be clearly distinguished from the noise level, although we tried to find a relatively small $\Delta \rho_{DM}$ value, since we did not want to diminish the local wake artificially. Therefore, the limits derived from this method should be considered as upper limits for the local wake extent and are hereafter called $R_{\rm local}$. In addition to this, we also calculated the theoretical first-order perturbation response of the DM halo according to the Chandrasekhar formalism \citep[see][]{Mulder_et_al_1983, Binney_and_Tremaine_2008} to the surface overdensity (Figure~\ref{fig:binney_tremaine_torques_point}). If we assume a homogeneous medium and a satellite on a straight trajectory, according to \cite{Mulder_et_al_1983} the first-order term in the perturbation calculation is given by

\begin{align}\label{eq:mulder}
   \rho - \rho_0  =& \frac{G M \rho_0}{r \sigma_0^2}{\rm e}^{-0.5 (|\vec{v}_{\rm sat}|/\sigma_0)^2 (1-\cos(\theta) ^2)} \nonumber \\
   & \cdot \left [ 1 - {\rm erfc} \left ( \frac{(|\vec{v}_{\rm sat}|/\sigma_0) \cos(\theta)}{\sqrt{2}} \right )\right ],
\end{align}

\noindent with $\rho_0$ denoting the initial (constant) density, $M$ the mass of the satellite, $r := |\vec{r}|$ the distance from the satellite, $\sigma_0$ the 1D velocity dispersion, calculated as the mean of the 3D velocity dispersion of all DM particles which are within a 30~kpc sphere around the satellite's center, $\vec{v}_{\rm sat}$ the satellite's velocity, and $\theta$ the angle between $\vec{v}_{\rm sat}$ and $\vec{r}$. After manipulating both sides, we obtain $R_{\alpha}$, which is the radial extent of the local wake given by theoretical prediction for a given overdensity $\Delta_\rho \equiv \frac{\rho}{ \rho_0}-1$:

\begin{align}\label{eq:overdensity}
    R_\alpha =& \frac{\alpha}{\Delta_\rho} \frac{G M }{\sigma_0^2}{\rm e}^{-0.5 (|\vec{v}_{\rm sat}|/\sigma_0)^2 (1-\cos(\theta) ^2)} \nonumber \\
   & \cdot \left [ 1 - {\rm erfc} \left ( \frac{(|\vec{v}_{\rm sat}|/\sigma_0) \cos(\theta)}{\sqrt{2}} \right )\right ] ,
\end{align}

\noindent where we included a dimensionless parameter $\alpha$ to match the theoretical predictions to our simulations. This is necessary, since our simulations do not obey the necessary conditions (an infinite and homogeneous medium) for Equation~\eqref{eq:mulder}. We find that a variable $\alpha$ in the range 5--8 yields good accordance with the data, see Figure~\ref{fig:halfcumulative-time-series}. It was impossible to find a single $\alpha$-value to match the entire data set and, in order to match the simulations, $\alpha$ had to decrease as time increased. Furthermore, this analysis, especially Equation~\eqref{eq:mulder}, is the reason for our choice of normalization in Figure~\ref{fig:binney_tremaine}.

In order to compare the calculated half-torque radius, $R_{50\%}$, to the classical prediction, we can now use Equation~\eqref{eq:overdensity} and estimate the radius of the overdensity from the contours as well as our visual approach from the simulation, which we call $R_{\rm local}$, for which we use $\Delta_\rho \geq 0.8$ and $\Delta_\rho \geq 1.0$ (see Figure~\ref{fig:binney_tremaine_torques_point}) in order to calculate the distance from the satellite's position to the furthest point on the contour of $\Delta_\rho$. 

Before the satellite sinks to the center (at $\sim$1.8~Gyr), there are three main regimes of its angular momentum evolution.  First, the angular momentum very slowly decreases between 0 and $\sim$0.55~Gyr; this is followed by a faster decay between $\sim$0.55~Gyr and $\sim$0.7~Gyr; then, it abruptly drops between $\sim$0.7~Gyr and $1.3$~Gyr (see Figures~\ref{fig:tot_torque} and \ref{fig:halfcumulative-time-series}). Those angular momentum loss regimes can be now also seen as the regimes where the $R_{75 \%}$ and then the $R_{50 \%}$ radii cross with the $R_{\rm local}$ radii. Since $R_{\rm local}$ dominates (before the first peri-center approach), it means that the majority of the torque comes from within the local wake radius and the satellite's decay is mainly dictated by the local wake description. Once $R_{50 \%}$ is larger than $R_{\rm local}$, which occurs roughly after the first peri-center, the bulk share of the torque on the satellite comes from regions outside of the local wake regime. The fact that we see this transition after the first peri-center is in good agreement with \citet{Weinberg_1986}.

As already mentioned in Section~\ref{sec:orbits}, we also investigated the possibility that we might have picked a particularly good initial setup to observe the modes and therefore we also simulated the purely radial $i=90^\circ$ case. This simulation shows a similar behaviour as that previously described (for more information on the $i = 90^\circ$ simulation, see Appendix~\ref{sec:90}).

\section{Discussion}\label{sec:Discussion}
\begin{figure*}[htb!]
\includegraphics[width=\textwidth]{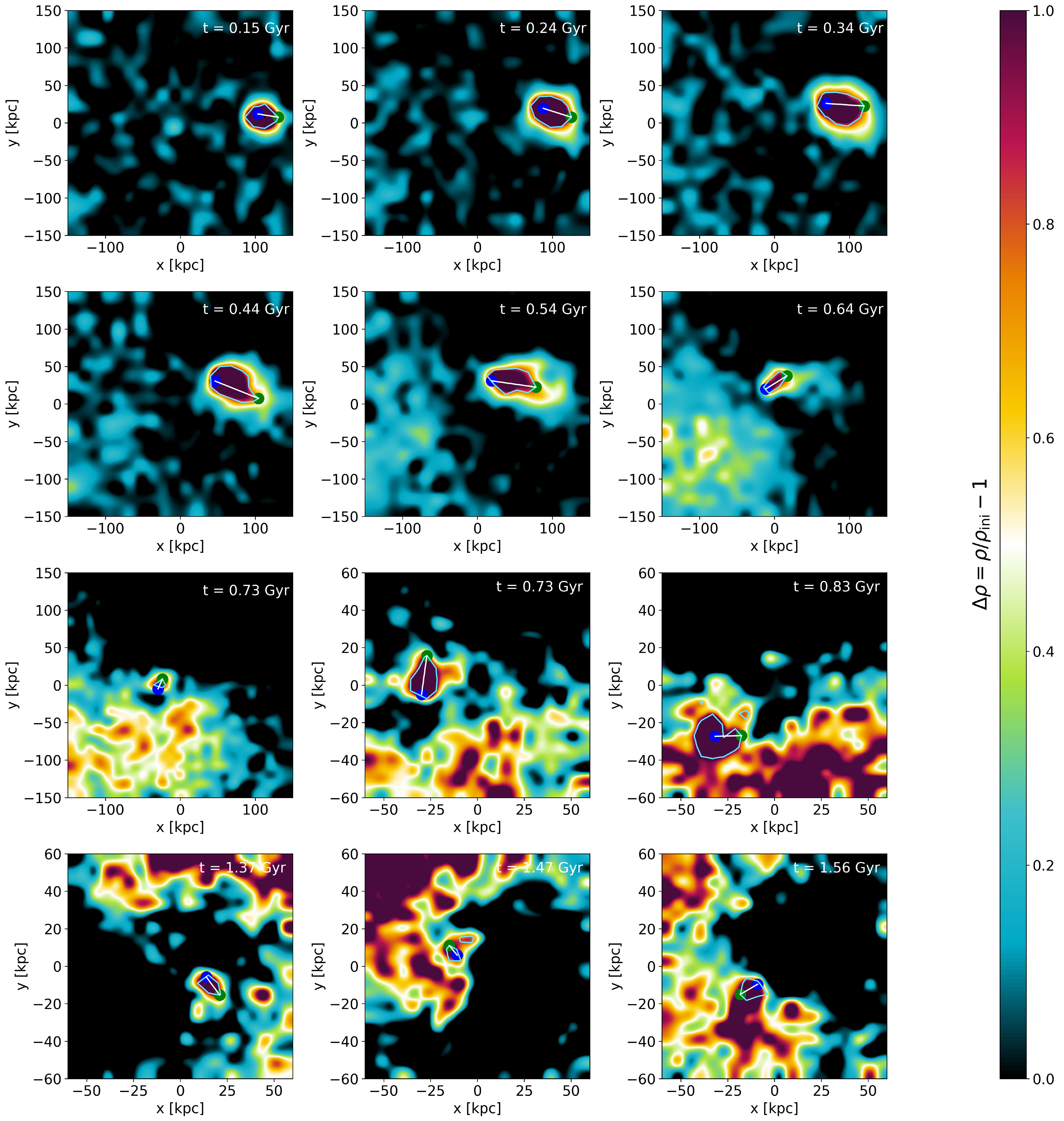}
\vspace{2pt}
\caption{This figure depicts the local wake in the HR $i=60^\circ$ rigid-satellite run and also shows how we estimated the local wake radius $R_{\rm local}$. The $\Delta \rho_{\rm DM}$ map has been smoothed by combining neighbouring cells and afterwards a contour plot with $\Delta \rho_{\rm DM} =0.8$ (for the time range 0--0.62~Gyr) or $\Delta \rho_{\rm DM}=1.0$ (for the time range 0.62--1.58~Gyr) has been superimposed. The blue circle represents the satellite's position and the green circle represents the point furthest away from the former. The straight white line indicates the extent of the local wake, which we use for the $R_{\rm local}$ values.}
\vspace{12pt}
\label{fig:binney_tremaine_torques_point}
\end{figure*}

In this work, we have found clear evidence that the halo develops a complex modal structure in its density field when responding to the perturbation induced by a massive satellite. Up to resolutions of the primary halo of a few million particles, namely comparable to what can be achieved today in zoom-in cosmological simulations of galaxy formation, and typical of past published $N$-body simulations of the galaxy-satellite interaction \citep[e.g.][]{Kazantzidis_et_al_2008, Kazantzidis_et_al_2009, Purcell_et_al_2009, Villalobos_et_al_2008, Villalobos_et_al_2009, Villalobos_et_al_2010, van_den_Bosch_et_al_2018, van_den_Bosch_and_Ogiya_2018}, only the lowest-order modes, the $\ell=1$ dipole and the $\ell=2$, which represents the stellar disk potential, can be clearly identified. Resolutions of 20 million particles or higher are necessary to clearly resolve higher-order modes, which might be responsible for the visible asymmetry in the density distribution (see Figures~\ref{fig:localwake}, \ref{fig:binney_tremaine}, and \ref{fig:binney_tremaine_torques_point}).

Our results have also shown how a significant amount of the net torque on the satellite comes from regions that are further away than the local wake. This is clear in the analysis of the rigid satellites' decay. The scale of the drag becomes comparable to the scale of the $\ell=1$ distortion soon after the first orbit, which can be also seen in Figure~\ref{fig:mollweide60} as the dipole across half the sky. This mode then appears to be mostly responsible for the orbital decay at subsequent orbits. Higher-order modes may contribute to a lesser extent, and indeed we found evidence that a visible $\ell=3$ distortion sets a relevant scale for the torque during an intermediate phase. The local wake requires modes which are at least $\ell=5$ and higher, although the correspondence is not obvious, and its role in the drag is only important during the first orbit, as originally argued by \cite{Weinberg_1989}. We note that the dominant role of the $\ell=1$ mode explains why the orbital decay curve does not depend on resolution; as shown by the power spectrum analysis, indeed, this mode is always sufficiently well resolved, even at the lowest resolution level. 

In order to strengthen the proposed relation between the observed spatially localized overdensity and the high $\ell$ features in the power spectrum, we artificially weakened the spatial overdensity trailing the satellite in our simulation and recalculated the power spectrum to check if specific $C_\ell$ features are damped. Therefore, we performed the following procedure:

\begin{itemize}

    \item In the HR rigid-satellite simulation we replace, in a single snapshot at time $t_{\rm sample}$, all particles within a sphere with radius $r_{\rm sphere}$ with particles from the initial conditions.
    
    \item If the sampled sphere, at time $t_{\rm sample}$, contains more particles than at time zero, we randomly shuffle the remaining particles within the sphere of radius $r_{\rm sphere}$.
    
    \item We perform the above procedure for each time step at 10 different radii: $r_{\rm sphere} =5$, 10, 15, 20, 25, 30, 35, 40, 45, and 50~kpc.
    
    \item For each $r_{\rm sphere}$, we recalculate the power spectrum.
    
\end{itemize}

\begin{figure*}[htb!]
\centering
\includegraphics[width=\textwidth]{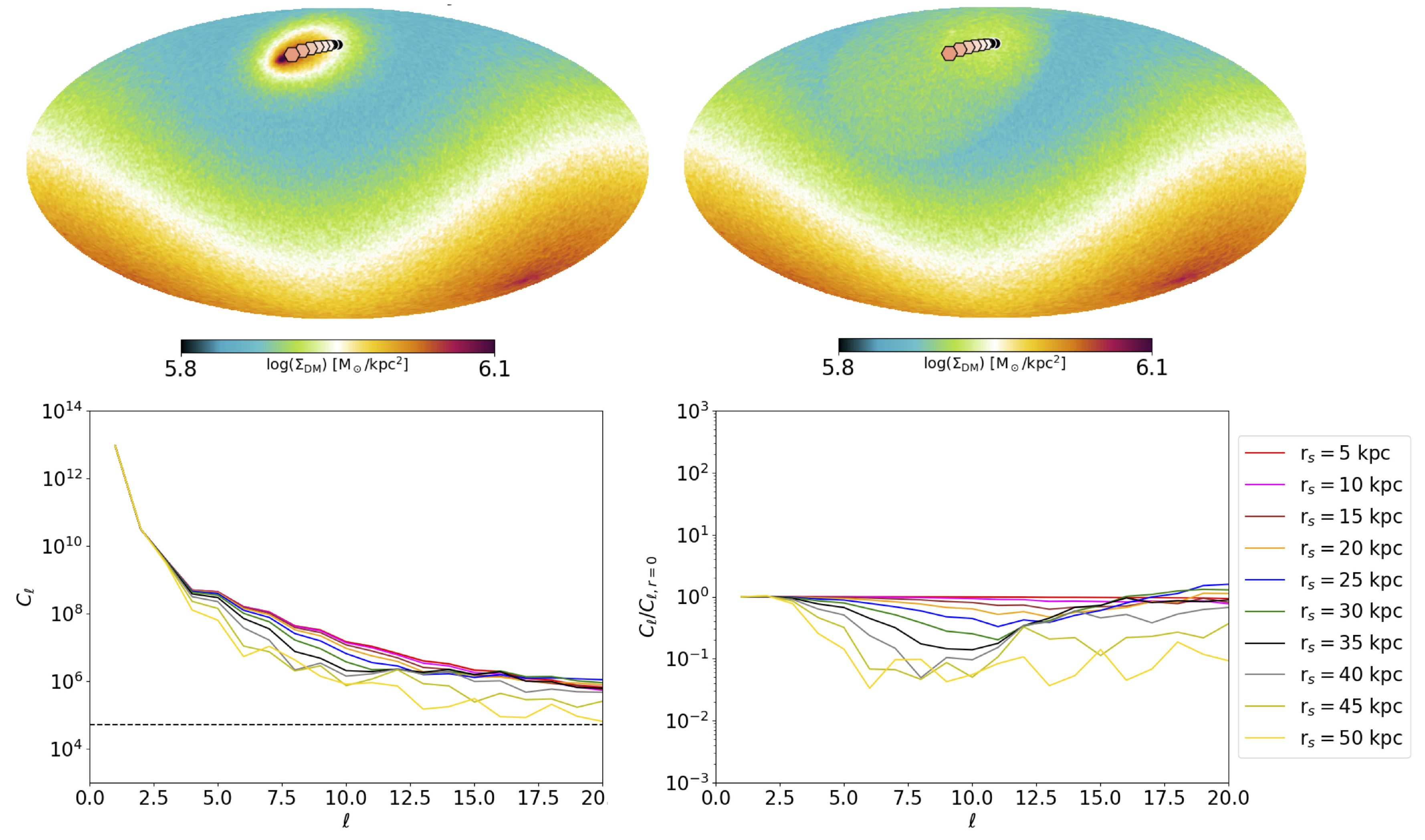}
\vspace{12pt}
\caption{We depict the smoothing procedure which we applied in order to weaken the trailing overdensity of the satellite. The \textit{top left} plot shows the healpix projection of the surface density $\Sigma_{\rm DM}$ of the MWs DM halo at time $t=0.39$~Gyr, which corresponds to the maximum of $R_\alpha$ and $R_{\rm local}$ (see Figure~\ref{fig:tot_torque}). On the \textit{top right}, we see the surface density after redistributing particles within a sphere of radius $r_{\rm sphere} = 50$~kpc around the rigid satellite's position. As a result, we erased the local overdensity in the surface density. The \textit{bottom} panels show the not normalized (\textit{left}) and normalized (\textit{right}) power spectra for 10 different radii $r_{\rm sphere} =  5$, 10, 15, 20, 25, 30, 35, 40, 45, and 50~kpc at time $t=0.39$~Gyr. Once we reach a radius between 40 and 50 kpc, we start to observe a dampening of the $\ell=4$ mode, which is roughly the mode we associate with the local wake (above $\ell=5$). Higher $\ell$ modes are affected by even smaller radii and therefore have to be attributed to features that are smaller than the local wake. Additionally, we see that the low-order $\ell$ modes are unaffected by altering the local neighborhood of the satellite.}
\vspace{12pt}
\label{fig:smoothed}
\end{figure*}

By sampling from the initial conditions, we minimize artificial over- and underdensities on the healpix maps, which would arise from simply randomly shuffling particles within the sphere of radius $r_{\rm sphere}$. Furthermore, we also conserve the sphere’s mass, which could not be done by simply sampling from the original density profile of the main MW halo. In Figure~\ref{fig:smoothed}, we present the results of this procedure at $t=0.39$~Gyr, which corresponds to the peak of $R_\alpha$ and $R_{\rm local}$. It is clear that we can actually smooth out the trailing overdensity with $r_{\rm sphere}$ being between 40 and 50~kpc. This leads to the damping of $\ell$ values larger than 4, which is roughly the mode that we previously assigned to the local wake. Furthermore, by doing this, we also obtain $\ell=1$ and $\ell=3$ modes that are unaffected by this and therefore are dominated by a large-scale response which is independent of local features. Remarkably, we also verified that the mode is prominently present irrespective of the procedure to center the primary halo when computing the power spectrum. While early work in the 1980's might have exaggerated the dipolar mode effect in simulations that had rigid halos pinned to the initial center of mass position, our HR simulations of completely live, self-consistent galaxy models demonstrate that this mode is physical (i.e. not of numerical origin) and very strong, being clearly discernible in density maps. The question then arises of why the global modes were not found previously in simulations, apart from the work of Weinberg and collaborators. One reason could simply be that, as we showed, it is only at very high resolution that the modes, including $\ell=1$, become visually clear in density maps of the primary halo. At lower resolution, their presence can only be determined by computing the power spectrum of density which, however, is not common practice in studies of satellite-halo interactions. Even in the latter case, it is only at high resolution, exceeding 10 million particles, that the discreteness noise level becomes low enough to resolve the correct mode amplitude. Hence the mode(s) might have been simply missed in the noisy $N$-body simulations. Not surprisingly, a low noise is a natural feature of SCF codes, which did spot the modes about three decades ago.

In order to challenge our findings, we investigated the possibility that differences in the gravity solver implementation, here specifically for tree-codes, might affect the global modes. Hence, we re-ran the ULR simulation with \textsc{Pkdgrav{1}} \citep[][]{Stadel_et_al_2001}, by switching off the hydrodynamics in \textsc{Gasoline2} \citep[][]{Wadsley_et_al_2004, Wadsley_et_al_2017}, but we did not find any differences with the \textsc{Pkdgrav{3}} simulation. Furthermore, we also switched off the octupole moment for the force calculation, which, again, yielded no deviation from our results in this work.

More quiet, stable galaxy initial conditions, such as those borne out of the Widrow \& Dubinski's initial condition generator that we adopted here, might also favour resolving global modes compared to past work adopting approximate initial condition generators which, for example, assumed a Maxwellian velocity distribution for NFW DM halos rather than using the velocity distribution associated with the corresponding density distribution \citep[e.g.][]{Kazantzidis_et_al_2004}.

Recently, \citet{Cunningham_et_al_2020} have shown that a merger of an LMC-like satellite with an MW-like galaxy can be observed as a large-scale $\ell=1$ mode in the spherical harmonics decomposition of the stellar halo velocities. A possible observation with Gaia via the halo stars is certainly possible, but they argue that cosmological simulations are needed in order to investigate the actual observability of such modes, since multiple mergers, stellar in-situ formation, and deviations in the DM halo from spherical symmetry may affect or even hide the signal. Considering the resolution analysis in this work, we can confirm that the resolution in currently ongoing cosmological simulations (and zoom-in simulations) is already high enough to reveal the DM response to an infalling satellite. Such an analysis will be part of future work.

We showed that the local wake is important at the first peri-center approach and we can attribute roughly $1/3$ of the total angular momentum loss to it. However, after the first peri-center, the local wake becomes less important. Eventually, it vanishes in terms of density contrast and, in fact, as we mentioned, it appears that after the first peri-center, most of the angular momentum is lost due to the global modes. This is surprising since previous works have found good agreement with \cite{Chandrasekhar_1943, Binney_and_Tremaine_2008}, which describes the force on a perturber via

\begin{equation}
    \frac{d \vec{v}_{\rm M}}{dt} = -16 \pi^2 (\ln{\Lambda}) G^2\frac{ m (M+m)}{|\vec{v}_{\rm M}|^3} \int_0^{|v_{\rm M}|} v^2 f(v) dv \vec{v}_{\rm M},
\end{equation}

\noindent with $m$ denoting the mass of the background particles, $M$ the mass of the perturber, $f(v)$ the velocity distribution of the background particles, $\vec{v}_{\rm M}$ the velocity of the perturber, and $\Lambda = b_{\rm max}/b_{\rm min}$, with $b_{\rm max}$ being the maximum impact parameter and $b_{\rm min}$ the 90$^\circ$ deflection angle. We calculated the expected decay $t_{\rm df}$ assuming a \citet{Hernquist_1990} profile, since this leads to a $t_{\rm df}$ formula which can be easily numerically integrated \citep[][]{SouzaLima_et_al_2017,Tamburello_et_al_2017}. In order to get this estimate, we approximated the NFW profile to a Hernquist profile \citep[][]{Springel_et_al_2005} and thus obtained $t_{\rm df} \approx 2.8$~Gyr. We should now note that the orbital decay of satellites in $N$-body simulations has been routinely matched with Chandrasekhar's formula by adjusting the value of the Coulomb logarithm, $\ln \Lambda$, in the force expression \citep[e.g.][]{van_den_Bosch_et_al_1999, Colpi_et_al_1999}. We could not find a very good match to our orbital decay by simply scaling the Coulomb logarithm, always ending with a timescale too long compared to what we measured in the simulations. However, nearly all works performing the orbital decay timescale analysis adopted satellites which had masses 20 to 50 times lower than the mass of the primary, whereas the satellite that we considered in this work is more massive. With a more massive satellite, halo mode excitation should be stronger, hence a larger departure from the conventional dynamical friction theory is expected. In the future, it would be interesting to perform a large set of high-resolution simulations with satellites of different masses. One cannot exclude that, as the satellite's mass is decreased, the local wake becomes proportionally more important for the drag relative to the global modes. Furthermore, previous numerical simulations have also shown that dynamical friction for binary black holes in constant-density cores is inefficient and leads to stalling, even before a hard binary is formed \citep[see][]{Goerdt_et_al_2006, Cole_et_al_2012, Petts_et_al_2016, DiCintio_et_al_2017, Tamfal_et_al_2018}. This has been always explained in the context of the local dynamical friction theory and therefore also here the question remains, how does the global mode change with a cored DM profile. Considering that DM modes might be also important in the BH binary formation, previous works that investigate the binary formation in cosmological simulations \citep[e.g.][]{Tremmel_et_al_2015,Habouzit_et_al_2017,Bellovary_et_al_2018,Bortolas_et_al_2020} might have already seen the effect of DM modes in their simulations. Such a study will be the subject of future work carried out by the authors.


\section{Concluding remarks}
In this work, we simulated multiple encounters of an LMC-like satellite with an MW-like galaxy. We investigated four different orbital inclinations as well as four different resolutions. We studied the origin of the gravitational drag that causes the satellite's decay, and compared the results of the analysis over more than three orders of magnitude in resolution. Our findings can be summarized as follows: 
\begin{itemize}

    \item The halo develops global modes in response to the satellite's perturbation, which are clearly identifiable in the density power spectrum.
    
    \item We can detect a strong $\ell = 1$ dipolar mode at every resolution.
    
    \item The global mode and the local wake are present irrespective of resolution, but the associated signal-to-noise ratio degrades significantly with resolution, so that at low and ultra-low resolution only the lowest-order, dipolar mode is clearly resolved. The higher the resolution, the better we can obtain higher-order $\ell$ modes.
    
    \item The behaviour of the satellite's orbital decay, as measured by the evolution of the distance from the center of mass of the bound particles to the center of the DM halo of the MW, is independent of resolution. This reflects the fact that the $\ell=1$ mode, which is the one that appears to produce the largest torque, as well as the $\ell=2$ mode, which is always the same since it depicts the stellar disk, are always resolved.
    
    \item We do not observe different results in the angular momentum loss depending on the resolution of the simulation, in agreement with the previous point.
    
    \item The largest angular momentum loss, after the first pericenter passage, comes from torques that originate from a region exterior to that in which the local wake develops.
    
\end{itemize}


\begin{acknowledgments}
We made use of the analysis code pynbody \citep[\url{https://github.com/pynbody/pynbody};][]{Pontzen_et_al_2013}. We want to thank the anonymous referee for the useful comments that helped us to further improve this paper. Furthermore, we also want to thank the organisers of the KITP program ``Dynamical Models for Stars and Gas in Galaxies in the Gaia Era'', Jo Bovy, James Bullock, Sukanya Chakrabarti, and Heidi Newberg, during which this work started. This research was supported in part by the National Science Foundation under Grant No. NSF PHY-1748958. Simulations were performed on the Piz Daint supercomputer of the Swiss National Supercomputing Centre (CSCS) under the project id s1014. PRC, LM and TT acknowledge support from the Swiss National Science Foundation under the grant 200020\_178949. AB acknowledges support from the  Natural Sciences and Engineering Research Council of Canada.\\
The data that support the findings of this study are available upon reasonable request from the authors.
\end{acknowledgments}


\bibliographystyle{aasjournal}
\bibliography{dm_wake}

\begin{thebibliography}{}
\expandafter\ifx\csname natexlab\endcsname\relax\def\natexlab#1{#1}\fi
\providecommand{\url}[1]{\href{#1}{#1}}
\providecommand{\dodoi}[1]{doi:~\href{http://doi.org/#1}{\nolinkurl{#1}}}
\providecommand{\doeprint}[1]{\href{http://ascl.net/#1}{\nolinkurl{http://ascl.net/#1}}}
\providecommand{\doarXiv}[1]{\href{https://arxiv.org/abs/#1}{\nolinkurl{https://arxiv.org/abs/#1}}}

\bibitem[{{Bekenstein}(1989)}]{Bekenstein_1989}
{Bekenstein}, J.~D. 1989, International Journal of Theoretical Physics, 28,
  967, \dodoi{10.1007/BF00670342}

\bibitem[{{Bellovary} {et~al.}(2019){Bellovary}, {Cleary}, {Munshi}, {Tremmel},
  {Christensen}, {Brooks}, \& {Quinn}}]{Bellovary_et_al_2018}
{Bellovary}, J.~M., {Cleary}, C.~E., {Munshi}, F., {et~al.} 2019, \mnras, 482,
  2913, \dodoi{10.1093/mnras/sty2842}

\bibitem[{{Benson}(2005)}]{Benson_2005}
{Benson}, A.~J. 2005, \mnras, 358, 551,
  \dodoi{10.1111/j.1365-2966.2005.08788.x}

\bibitem[{{Benson} {et~al.}(2002){Benson}, {Lacey}, {Baugh}, {Cole}, \&
  {Frenk}}]{Benson_et_al_2002}
{Benson}, A.~J., {Lacey}, C.~G., {Baugh}, C.~M., {Cole}, S., \& {Frenk}, C.~S.
  2002, \mnras, 333, 156, \dodoi{10.1046/j.1365-8711.2002.05387.x}

\bibitem[{{Binney} \& {Tremaine}(2008)}]{Binney_and_Tremaine_2008}
{Binney}, J., \& {Tremaine}, S. 2008, {Galactic Dynamics: Second Edition}

\bibitem[{{Blumenthal} {et~al.}(1984){Blumenthal}, {Faber}, {Primack}, \&
  {Rees}}]{Blumenthal_et_al_1984}
{Blumenthal}, G.~R., {Faber}, S.~M., {Primack}, J.~R., \& {Rees}, M.~J. 1984,
  \nat, 311, 517, \dodoi{10.1038/311517a0}

\bibitem[{{Bortolas} {et~al.}(2020){Bortolas}, {Capelo}, {Zana}, {Mayer},
  {Bonetti}, {Dotti}, {Davies}, \& {Madau}}]{Bortolas_et_al_2020}
{Bortolas}, E., {Capelo}, P.~R., {Zana}, T., {et~al.} 2020, \mnras, 498, 3601,
  \dodoi{10.1093/mnras/staa2628}

\bibitem[{{Bullock} {et~al.}(2001){Bullock}, {Kolatt}, {Sigad}, {Somerville},
  {Kravtsov}, {Klypin}, {Primack}, \& {Dekel}}]{Bullock_et_al_2001}
{Bullock}, J.~S., {Kolatt}, T.~S., {Sigad}, Y., {et~al.} 2001, \mnras, 321,
  559, \dodoi{10.1046/j.1365-8711.2001.04068.x}

\bibitem[{{Capaccioli}(1989)}]{Capaccioli_1989}
{Capaccioli}, M. 1989, in World of Galaxies (Le Monde des Galaxies), ed.
  J.~{Corwin}, Harold~G. \& L.~{Bottinelli}, 208--227

\bibitem[{{Capelo} {et~al.}(2015){Capelo}, {Volonteri}, {Dotti}, {Bellovary},
  {Mayer}, \& {Governato}}]{Capelo_et_al_2015}
{Capelo}, P.~R., {Volonteri}, M., {Dotti}, M., {et~al.} 2015, \mnras, 447,
  2123, \dodoi{10.1093/mnras/stu2500}

\bibitem[{{Chandrasekhar}(1943)}]{Chandrasekhar_1943}
{Chandrasekhar}, S. 1943, \apj, 97, 255, \dodoi{10.1086/144517}

\bibitem[{{Choi} {et~al.}(2009){Choi}, {Weinberg}, \& {Katz}}]{Choi_et_al_2009}
{Choi}, J.-H., {Weinberg}, M.~D., \& {Katz}, N. 2009, \mnras, 400, 1247,
  \dodoi{10.1111/j.1365-2966.2009.15556.x}

\bibitem[{{Cole} {et~al.}(2012){Cole}, {Dehnen}, {Read}, \&
  {Wilkinson}}]{Cole_et_al_2012}
{Cole}, D.~R., {Dehnen}, W., {Read}, J.~I., \& {Wilkinson}, M.~I. 2012, \mnras,
  426, 601, \dodoi{10.1111/j.1365-2966.2012.21885.x}

\bibitem[{{Colpi} {et~al.}(1999){Colpi}, {Mayer}, \&
  {Governato}}]{Colpi_et_al_1999}
{Colpi}, M., {Mayer}, L., \& {Governato}, F. 1999, \apj, 525, 720,
  \dodoi{10.1086/307952}

\bibitem[{{Colpi} \& {Pallavicini}(1998)}]{Colpi_et_al_1998}
{Colpi}, M., \& {Pallavicini}, A. 1998, \apj, 502, 150, \dodoi{10.1086/305877}

\bibitem[{{Cunningham} {et~al.}(2020){Cunningham}, {Garavito-Camargo},
  {Deason}, {Johnston}, {Erkal}, {Laporte}, {Besla}, {Luger}, \&
  {Sanderson}}]{Cunningham_et_al_2020}
{Cunningham}, E.~C., {Garavito-Camargo}, N., {Deason}, A.~J., {et~al.} 2020,
  \apj, 898, 4, \dodoi{10.3847/1538-4357/ab9b88}

\bibitem[{{Dehnen}(2001)}]{Dehnen_2001}
{Dehnen}, W. 2001, \mnras, 324, 273, \dodoi{10.1046/j.1365-8711.2001.04237.x}

\bibitem[{{Di Cintio} {et~al.}(2017){Di Cintio}, {Tremmel}, {Governato},
  {Pontzen}, {Zavala}, {Bastidas Fry}, {Brooks}, \&
  {Vogelsberger}}]{DiCintio_et_al_2017}
{Di Cintio}, A., {Tremmel}, M., {Governato}, F., {et~al.} 2017, \mnras, 469,
  2845, \dodoi{10.1093/mnras/stx1043}

\bibitem[{{Dubinski} {et~al.}(2009){Dubinski}, {Berentzen}, \&
  {Shlosman}}]{Dubinski_et_al_2009}
{Dubinski}, J., {Berentzen}, I., \& {Shlosman}, I. 2009, \apj, 697, 293,
  \dodoi{10.1088/0004-637X/697/1/293}

\bibitem[{{Fakhouri} {et~al.}(2010){Fakhouri}, {Ma}, \&
  {Boylan-Kolchin}}]{Fakhouri_et_al_2010}
{Fakhouri}, O., {Ma}, C.-P., \& {Boylan-Kolchin}, M. 2010, \mnras, 406, 2267,
  \dodoi{10.1111/j.1365-2966.2010.16859.x}

\bibitem[{{Garavito-Camargo} {et~al.}(2019){Garavito-Camargo}, {Besla},
  {Laporte}, {Johnston}, {G{\'o}mez}, \& {Watkins}}]{Garavito_et_al_2019}
{Garavito-Camargo}, N., {Besla}, G., {Laporte}, C. F.~P., {et~al.} 2019, \apj,
  884, 51, \dodoi{10.3847/1538-4357/ab32eb}

\bibitem[{{Goerdt} {et~al.}(2006){Goerdt}, {Moore}, {Read}, {Stadel}, \&
  {Zemp}}]{Goerdt_et_al_2006}
{Goerdt}, T., {Moore}, B., {Read}, J.~I., {Stadel}, J., \& {Zemp}, M. 2006,
  \mnras, 368, 1073, \dodoi{10.1111/j.1365-2966.2006.10182.x}

\bibitem[{{G{\'o}rski} {et~al.}(2005){G{\'o}rski}, {Hivon}, {Banday},
  {Wandelt}, {Hansen}, {Reinecke}, \& {Bartelmann}}]{healpy1}
{G{\'o}rski}, K.~M., {Hivon}, E., {Banday}, A.~J., {et~al.} 2005, \apj, 622,
  759, \dodoi{10.1086/427976}

\bibitem[{{Habouzit} {et~al.}(2017){Habouzit}, {Volonteri}, \&
  {Dubois}}]{Habouzit_et_al_2017}
{Habouzit}, M., {Volonteri}, M., \& {Dubois}, Y. 2017, \mnras, 468, 3935,
  \dodoi{10.1093/mnras/stx666}

\bibitem[{{Helmi} {et~al.}(2018){Helmi}, {Babusiaux}, {Koppelman}, {Massari},
  {Veljanoski}, \& {Brown}}]{Helmi_et_al_2018}
{Helmi}, A., {Babusiaux}, C., {Koppelman}, H.~H., {et~al.} 2018, \nat, 563, 85,
  \dodoi{10.1038/s41586-018-0625-x}

\bibitem[{{Hernquist}(1990)}]{Hernquist_1990}
{Hernquist}, L. 1990, \apj, 356, 359, \dodoi{10.1086/168845}

\bibitem[{{Hernquist} \& {Ostriker}(1992)}]{Hernquist_et_al_1992}
{Hernquist}, L., \& {Ostriker}, J.~P. 1992, \apj, 386, 375,
  \dodoi{10.1086/171025}

\bibitem[{{Hernquist} {et~al.}(1995){Hernquist}, {Sigurdsson}, \&
  {Bryan}}]{Hernquist_et_al_1995}
{Hernquist}, L., {Sigurdsson}, S., \& {Bryan}, G.~L. 1995, \apj, 446, 717,
  \dodoi{10.1086/175829}

\bibitem[{{Hinshaw} {et~al.}(2013){Hinshaw}, {Larson}, {Komatsu}, {Spergel},
  {Bennett}, {Dunkley}, {Nolta}, {Halpern}, {Hill}, {Odegard}, {Page}, {Smith},
  {Weiland}, {Gold}, {Jarosik}, {Kogut}, {Limon}, {Meyer}, {Tucker}, {Wollack},
  \& {Wright}}]{Hinshaw_et_al_2013}
{Hinshaw}, G., {Larson}, D., {Komatsu}, E., {et~al.} 2013, \apjs, 208, 19,
  \dodoi{10.1088/0067-0049/208/2/19}

\bibitem[{{Hopkins} {et~al.}(2018){Hopkins}, {Wetzel}, {Kere{\v{s}}},
  {Faucher-Gigu{\`e}re}, {Quataert}, {Boylan-Kolchin}, {Murray}, {Hayward},
  {Garrison-Kimmel}, {Hummels}, {Feldmann}, {Torrey}, {Ma},
  {Angl{\'e}s-Alc{\'a}zar}, {Su}, {Orr}, {Schmitz}, {Escala}, {Sanderson},
  {Grudi{\'c}}, {Hafen}, {Kim}, {Fitts}, {Bullock}, {Wheeler}, {Chan},
  {Elbert}, \& {Narayanan}}]{Hopkins_et_al_2018}
{Hopkins}, P.~F., {Wetzel}, A., {Kere{\v{s}}}, D., {et~al.} 2018, \mnras, 480,
  800, \dodoi{10.1093/mnras/sty1690}

\bibitem[{{Kazantzidis} {et~al.}(2008){Kazantzidis}, {Bullock}, {Zentner},
  {Kravtsov}, \& {Moustakas}}]{Kazantzidis_et_al_2008}
{Kazantzidis}, S., {Bullock}, J.~S., {Zentner}, A.~R., {Kravtsov}, A.~V., \&
  {Moustakas}, L.~A. 2008, \apj, 688, 254, \dodoi{10.1086/591958}

\bibitem[{{Kazantzidis} {et~al.}(2004){Kazantzidis}, {Magorrian}, \&
  {Moore}}]{Kazantzidis_et_al_2004}
{Kazantzidis}, S., {Magorrian}, J., \& {Moore}, B. 2004, \apj, 601, 37,
  \dodoi{10.1086/380192}

\bibitem[{{Kazantzidis} {et~al.}(2009){Kazantzidis}, {Zentner}, {Kravtsov},
  {Bullock}, \& {Debattista}}]{Kazantzidis_et_al_2009}
{Kazantzidis}, S., {Zentner}, A.~R., {Kravtsov}, A.~V., {Bullock}, J.~S., \&
  {Debattista}, V.~P. 2009, \apj, 700, 1896,
  \dodoi{10.1088/0004-637X/700/2/1896}

\bibitem[{{Kazantzidis} {et~al.}(2005){Kazantzidis}, {Mayer}, {Colpi}, {Madau},
  {Debattista}, {Wadsley}, {Stadel}, {Quinn}, \&
  {Moore}}]{Kazantzidis_et_al_2005}
{Kazantzidis}, S., {Mayer}, L., {Colpi}, M., {et~al.} 2005, \apjl, 623, L67,
  \dodoi{10.1086/430139}

\bibitem[{{Kuhlen} {et~al.}(2012){Kuhlen}, {Vogelsberger}, \&
  {Angulo}}]{Kuhlen_et_al_2012}
{Kuhlen}, M., {Vogelsberger}, M., \& {Angulo}, R. 2012, Physics of the Dark
  Universe, 1, 50, \dodoi{10.1016/j.dark.2012.10.002}

\bibitem[{{Kuijken} \& {Dubinski}(1995)}]{Kuijken_Dubinski_1995}
{Kuijken}, K., \& {Dubinski}, J. 1995, \mnras, 277, 1341,
  \dodoi{10.1093/mnras/277.4.1341}

\bibitem[{{Lacey} \& {Cole}(1993)}]{Lacey_et_al_1993}
{Lacey}, C., \& {Cole}, S. 1993, \mnras, 262, 627,
  \dodoi{10.1093/mnras/262.3.627}

\bibitem[{{Li} \& {White}(2009)}]{Li_et_al_2009}
{Li}, C., \& {White}, S. D.~M. 2009, \mnras, 398, 2177,
  \dodoi{10.1111/j.1365-2966.2009.15268.x}

\bibitem[{{M{\'a}rquez} {et~al.}(2000){M{\'a}rquez}, {Lima Neto}, {Capelato},
  {Durret}, \& {Gerbal}}]{Marquez_et_al_2000}
{M{\'a}rquez}, I., {Lima Neto}, G.~B., {Capelato}, H., {Durret}, F., \&
  {Gerbal}, D. 2000, \aap, 353, 873.
\newblock \doarXiv{astro-ph/9911464}

\bibitem[{{Mollweide}(1805)}]{Mollweide_1805}
{Mollweide}, C.~B. 1805, {Mappirungskunst des Claudius Prolemaeus, ein Beytrag
  zur Geschichte der Landkarten}, ed. F.~{Zach}

\bibitem[{{Mulder}(1983)}]{Mulder_et_al_1983}
{Mulder}, W.~A. 1983, \aap, 117, 9

\bibitem[{{Navarro} {et~al.}(1996){Navarro}, {Frenk}, \&
  {White}}]{Navarro_et_al_1996}
{Navarro}, J.~F., {Frenk}, C.~S., \& {White}, S.~D.~M. 1996, \apj, 462, 563,
  \dodoi{10.1086/177173}

\bibitem[{{Ogiya} \& {Burkert}(2016)}]{Ogiya_et_al_2016}
{Ogiya}, G., \& {Burkert}, A. 2016, \mnras, 457, 2164,
  \dodoi{10.1093/mnras/stw091}

\bibitem[{{O'Leary} {et~al.}(2021){O'Leary}, {Moster}, {Naab}, \&
  {Somerville}}]{OLeary_et_al_2020}
{O'Leary}, J.~A., {Moster}, B.~P., {Naab}, T., \& {Somerville}, R.~S. 2021,
  \mnras, 501, 3215, \dodoi{10.1093/mnras/staa3746}

\bibitem[{{Petts} {et~al.}(2016){Petts}, {Read}, \&
  {Gualandris}}]{Petts_et_al_2016}
{Petts}, J.~A., {Read}, J.~I., \& {Gualandris}, A. 2016, \mnras, 463, 858,
  \dodoi{10.1093/mnras/stw2011}

\bibitem[{{Pontzen} {et~al.}(2013){Pontzen}, {Ro{\v{s}}kar}, {Stinson}, \&
  {Woods}}]{Pontzen_et_al_2013}
{Pontzen}, A., {Ro{\v{s}}kar}, R., {Stinson}, G., \& {Woods}, R. 2013,
  {pynbody: N-Body/SPH analysis for python}.
\newblock \doeprint{1305.002}

\bibitem[{{Potter} {et~al.}(2017){Potter}, {Stadel}, \&
  {Teyssier}}]{Potter_et_al_2017}
{Potter}, D., {Stadel}, J., \& {Teyssier}, R. 2017, Computational Astrophysics
  and Cosmology, 4, 2, \dodoi{10.1186/s40668-017-0021-1}

\bibitem[{{Power} {et~al.}(2003){Power}, {Navarro}, {Jenkins}, {Frenk},
  {White}, {Springel}, {Stadel}, \& {Quinn}}]{Power_et_al_2003}
{Power}, C., {Navarro}, J.~F., {Jenkins}, A., {et~al.} 2003, \mnras, 338, 14,
  \dodoi{10.1046/j.1365-8711.2003.05925.x}

\bibitem[{{Prugniel} \& {Simien}(1997)}]{Prugniel_Simien_1997}
{Prugniel}, P., \& {Simien}, F. 1997, \aap, 321, 111

\bibitem[{{Purcell} {et~al.}(2009){Purcell}, {Kazantzidis}, \&
  {Bullock}}]{Purcell_et_al_2009}
{Purcell}, C.~W., {Kazantzidis}, S., \& {Bullock}, J.~S. 2009, \apjl, 694, L98,
  \dodoi{10.1088/0004-637X/694/2/L98}

\bibitem[{{S{\'e}rsic}(1963)}]{Sersic_1963}
{S{\'e}rsic}, J.~L. 1963, Boletin de la Asociacion Argentina de Astronomia La
  Plata Argentina, 6, 41

\bibitem[{{S{\'e}rsic}(1968)}]{Sersic_1968}
---. 1968, {Atlas de Galaxias Australes}

\bibitem[{{Soko{\l}owska} {et~al.}(2017){Soko{\l}owska}, {Capelo}, {Fall},
  {Mayer}, {Shen}, \& {Bonoli}}]{Sokolowska_et_al_2017}
{Soko{\l}owska}, A., {Capelo}, P.~R., {Fall}, S.~M., {et~al.} 2017, \apj, 835,
  289, \dodoi{10.3847/1538-4357/835/2/289}

\bibitem[{{Souza Lima} {et~al.}(2017){Souza Lima}, {Mayer}, {Capelo}, \&
  {Bellovary}}]{SouzaLima_et_al_2017}
{Souza Lima}, R., {Mayer}, L., {Capelo}, P.~R., \& {Bellovary}, J.~M. 2017,
  \apj, 838, 13, \dodoi{10.3847/1538-4357/aa5d19}

\bibitem[{{Spitzer}(1942)}]{Spitzer_1942}
{Spitzer}, Lyman, J. 1942, \apj, 95, 329, \dodoi{10.1086/144407}

\bibitem[{{Springel} {et~al.}(2005){Springel}, {Di Matteo}, \&
  {Hernquist}}]{Springel_et_al_2005}
{Springel}, V., {Di Matteo}, T., \& {Hernquist}, L. 2005, \mnras, 361, 776,
  \dodoi{10.1111/j.1365-2966.2005.09238.x}

\bibitem[{{Stadel}(2001)}]{Stadel_et_al_2001}
{Stadel}, J.~G. 2001, PhD thesis, UNIVERSITY OF WASHINGTON

\bibitem[{{Taffoni} {et~al.}(2002){Taffoni}, {Mayer}, {Colpi}, \&
  {Governato}}]{Taffoni_et_al_2002}
{Taffoni}, G., {Mayer}, L., {Colpi}, M., \& {Governato}, F. 2002, Astronomical
  Society of the Pacific Conference Series, Vol. 253, {Disruption of Satellites
  in Cosmological Haloes.}, ed. R.~{Fusco-Femiano} \& F.~{Matteucci}, 273

\bibitem[{{Tamburello} {et~al.}(2017){Tamburello}, {Capelo}, {Mayer},
  {Bellovary}, \& {Wadsley}}]{Tamburello_et_al_2017}
{Tamburello}, V., {Capelo}, P.~R., {Mayer}, L., {Bellovary}, J.~M., \&
  {Wadsley}, J.~W. 2017, \mnras, 464, 2952, \dodoi{10.1093/mnras/stw2561}

\bibitem[{{Tamfal} {et~al.}(2018){Tamfal}, {Capelo}, {Kazantzidis}, {Mayer},
  {Potter}, {Stadel}, \& {Widrow}}]{Tamfal_et_al_2018}
{Tamfal}, T., {Capelo}, P.~R., {Kazantzidis}, S., {et~al.} 2018, \apjl, 864,
  L19, \dodoi{10.3847/2041-8213/aada4b}

\bibitem[{{Taylor} \& {Babul}(2001)}]{Taylor_et_al_2001}
{Taylor}, J.~E., \& {Babul}, A. 2001, \apj, 559, 716, \dodoi{10.1086/322276}

\bibitem[{{Terzi{\'c}} \& {Graham}(2005)}]{Terzic_Graham_2005}
{Terzi{\'c}}, B., \& {Graham}, A.~W. 2005, \mnras, 362, 197,
  \dodoi{10.1111/j.1365-2966.2005.09269.x}

\bibitem[{{Tremaine} \& {Weinberg}(1984)}]{Tremaine_et_al_1984}
{Tremaine}, S., \& {Weinberg}, M.~D. 1984, \mnras, 209, 729,
  \dodoi{10.1093/mnras/209.4.729}

\bibitem[{{Tremmel} {et~al.}(2015){Tremmel}, {Governato}, {Volonteri}, \&
  {Quinn}}]{Tremmel_et_al_2015}
{Tremmel}, M., {Governato}, F., {Volonteri}, M., \& {Quinn}, T.~R. 2015,
  \mnras, 451, 1868, \dodoi{10.1093/mnras/stv1060}

\bibitem[{{van den Bosch} {et~al.}(1999){van den Bosch}, {Lewis}, {Lake}, \&
  {Stadel}}]{van_den_Bosch_et_al_1999}
{van den Bosch}, F.~C., {Lewis}, G.~F., {Lake}, G., \& {Stadel}, J. 1999, \apj,
  515, 50, \dodoi{10.1086/307023}

\bibitem[{{van den Bosch} \& {Ogiya}(2018)}]{van_den_Bosch_and_Ogiya_2018}
{van den Bosch}, F.~C., \& {Ogiya}, G. 2018, \mnras, 475, 4066,
  \dodoi{10.1093/mnras/sty084}

\bibitem[{{van den Bosch} {et~al.}(2018){van den Bosch}, {Ogiya}, {Hahn}, \&
  {Burkert}}]{van_den_Bosch_et_al_2018}
{van den Bosch}, F.~C., {Ogiya}, G., {Hahn}, O., \& {Burkert}, A. 2018, \mnras,
  474, 3043, \dodoi{10.1093/mnras/stx2956}

\bibitem[{{Van Wassenhove} {et~al.}(2014){Van Wassenhove}, {Capelo},
  {Volonteri}, {Dotti}, {Bellovary}, {Mayer}, \&
  {Governato}}]{VanWassenhove_et_al_2014}
{Van Wassenhove}, S., {Capelo}, P.~R., {Volonteri}, M., {et~al.} 2014, \mnras,
  439, 474, \dodoi{10.1093/mnras/stu024}

\bibitem[{{Villalobos} \& {Helmi}(2008)}]{Villalobos_et_al_2008}
{Villalobos}, {\'A}., \& {Helmi}, A. 2008, \mnras, 391, 1806,
  \dodoi{10.1111/j.1365-2966.2008.13979.x}

\bibitem[{{Villalobos} \& {Helmi}(2009)}]{Villalobos_et_al_2009}
---. 2009, \mnras, 399, 166, \dodoi{10.1111/j.1365-2966.2009.15085.x}

\bibitem[{{Villalobos} {et~al.}(2010){Villalobos}, {Kazantzidis}, \&
  {Helmi}}]{Villalobos_et_al_2010}
{Villalobos}, {\'A}., {Kazantzidis}, S., \& {Helmi}, A. 2010, \apj, 718, 314,
  \dodoi{10.1088/0004-637X/718/1/314}

\bibitem[{{Wadsley} {et~al.}(2017){Wadsley}, {Keller}, \&
  {Quinn}}]{Wadsley_et_al_2017}
{Wadsley}, J.~W., {Keller}, B.~W., \& {Quinn}, T.~R. 2017, \mnras, 471, 2357,
  \dodoi{10.1093/mnras/stx1643}

\bibitem[{{Wadsley} {et~al.}(2004){Wadsley}, {Stadel}, \&
  {Quinn}}]{Wadsley_et_al_2004}
{Wadsley}, J.~W., {Stadel}, J., \& {Quinn}, T. 2004, \na, 9, 137,
  \dodoi{10.1016/j.newast.2003.08.004}

\bibitem[{{Weinberg}(1986)}]{Weinberg_1986}
{Weinberg}, M.~D. 1986, \apj, 300, 93, \dodoi{10.1086/163785}

\bibitem[{{Weinberg}(1989)}]{Weinberg_1989}
---. 1989, \mnras, 239, 549, \dodoi{10.1093/mnras/239.2.549}

\bibitem[{{Weinberg} \& {Katz}(2002)}]{Weinberg_et_al_2002}
{Weinberg}, M.~D., \& {Katz}, N. 2002, \apj, 580, 627, \dodoi{10.1086/343847}

\bibitem[{{Wetzel} {et~al.}(2016){Wetzel}, {Hopkins}, {Kim},
  {Faucher-Gigu{\`e}re}, {Kere{\v{s}}}, \& {Quataert}}]{Wetzel_et_al_2016}
{Wetzel}, A.~R., {Hopkins}, P.~F., {Kim}, J.-h., {et~al.} 2016, \apjl, 827,
  L23, \dodoi{10.3847/2041-8205/827/2/L23}

\bibitem[{{White}(1983)}]{White_1983}
{White}, S.~D.~M. 1983, \apj, 274, 53, \dodoi{10.1086/161425}

\bibitem[{{Widrow} \& {Dubinski}(2005)}]{Widrow_Dubinski_2005}
{Widrow}, L.~M., \& {Dubinski}, J. 2005, \apj, 631, 838, \dodoi{10.1086/432710}

\bibitem[{{Widrow} {et~al.}(2008){Widrow}, {Pym}, \&
  {Dubinski}}]{Widrow_et_al_2008}
{Widrow}, L.~M., {Pym}, B., \& {Dubinski}, J. 2008, \apj, 679, 1239,
  \dodoi{10.1086/587636}

\bibitem[{Zonca {et~al.}(2019)Zonca, Singer, Lenz, Reinecke, Rosset, Hivon, \&
  Gorski}]{healpy2}
Zonca, A., Singer, L., Lenz, D., {et~al.} 2019, Journal of Open Source
  Software, 4, 1298, \dodoi{10.21105/joss.01298}

\end{thebibliography}

\newpage

\appendix

\section{90-degree live-satellite runs}\label{sec:90}

\begin{figure*}[htb!]
\centering
\includegraphics[width=0.7\textwidth]{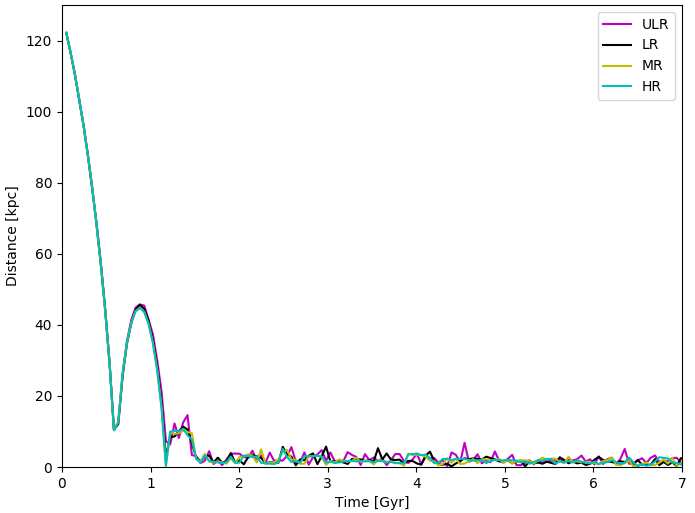}
\vspace{2pt}
\caption{Distance as a function of time between the center of mass of the bound live-satellite particles and the center of mass of the MW halo, for the $90^\circ$ run in all four resolutions. Again, all curves depict very similar, although slightly different, orbital histories.}
\vspace{12pt}
\label{fig:appendix_90}
\end{figure*}

\begin{figure*}[htb!]
\includegraphics[width=1.0\textwidth]{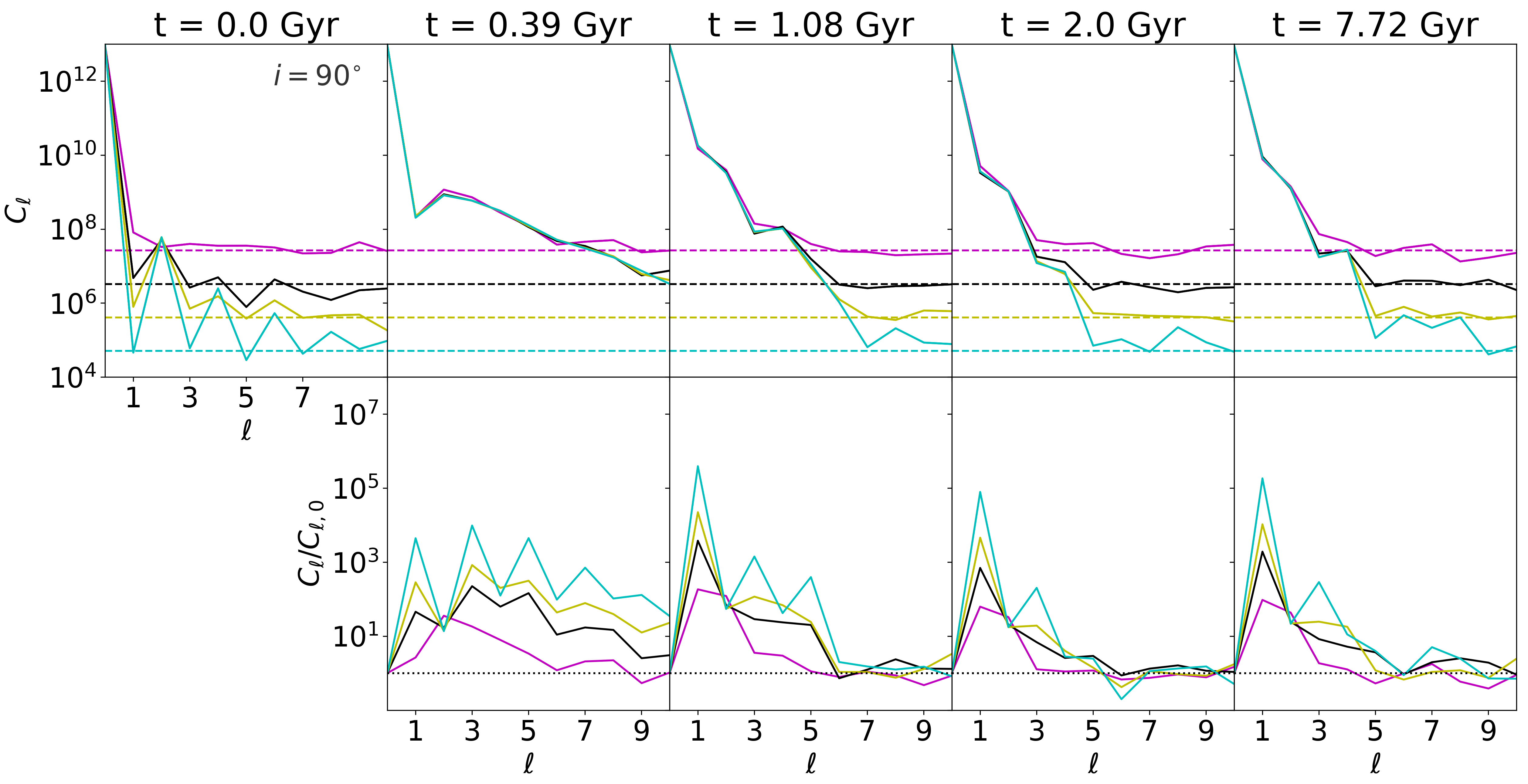}
\vspace{2pt}
\caption{This figure shows the power spectrum of the $90^\circ$ live-satellite runs (HR, MR, LR, and ULR). The figure is grouped similarly as Figure~\ref{fig:powerspectrum_norm}: the first row shows the (not normalized) power spectrum of the DM MW halo, with the first column showing the initial power spectrum and the horizontal dashed lines showing the theoretical sampling noise of the DM halo. The columns show the evolution of the power spectrum at the same times as in Figure~\ref{fig:powerspectrum_norm} (from left to right: $t= 0.39$, 1.08, 2.0, and 7.72~Gyr). Despite having different orbital parameters, the power spectrum shows a similar behaviour as our simulations with $i=0^\circ, 30^\circ$, and $60^\circ$.}
\vspace{12pt}
\label{fig:appendix_powerspectrum_norm}
\end{figure*}

In this section, we present the results of our $90^\circ$ live-satellite simulations. In Figure~\ref{fig:appendix_90}, we show the distance between the center of the bound particles of the satellite and the COM of the DM halo of the MW, similarly to what was shown in Figure~\ref{fig:distance} for the $60^\circ$ live-satellite simulations. In Figure~\ref{fig:appendix_powerspectrum_norm}, we also present the power spectrum of these runs, similarly to what was shown in Figure~\ref{fig:powerspectrum_norm} for the $0^\circ$, $30^\circ$, and $60^\circ$ live-satellite simulations. Despite the completely different orbit, in which the satellite sinks much faster, we clearly see the same excitation of low-order $\ell$ modes and the same negligible dependence on resolution. Therefore, we argue that our results are robust and not strongly affected by the orbital parameters.


\section{Relaxation of the HR MW run}\label{sec:RELAX}
\begin{figure*}[htb!]
\includegraphics[width=1.0\textwidth]{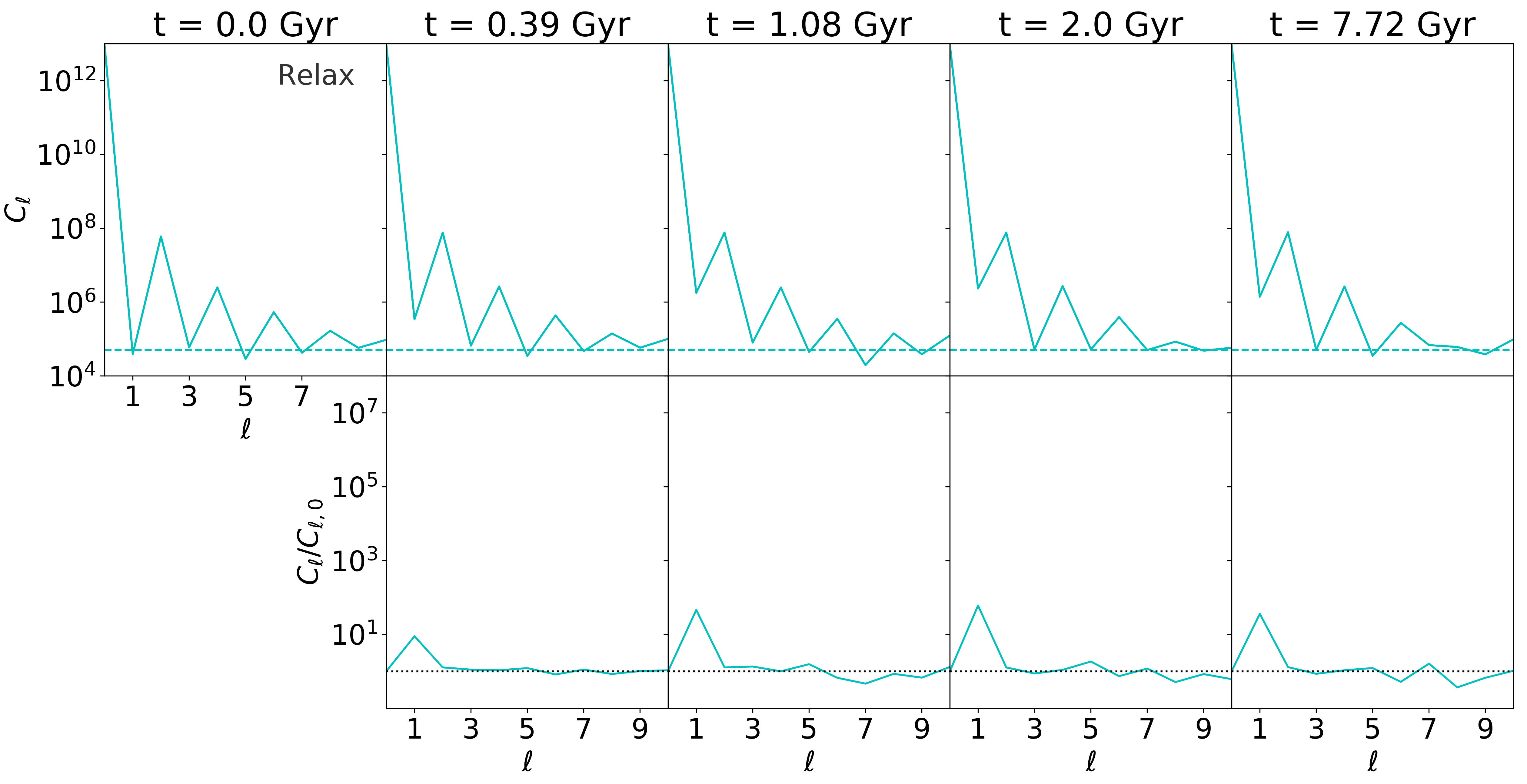}
\vspace{2pt}
\caption{The figure is grouped similarly as Figure~\ref{fig:powerspectrum_norm}: the first row shows the (not normalized) power spectrum of the DM MW halo, with the first column showing the initial power spectrum and the horizontal dashed lines showing the theoretical sampling noise of the DM halo. The columns show the evolution of the power spectrum at the same times: $t= 0.39$, 1.08, 2.0, and 7.72~Gyr. 
Since we constructed our initial models with a self-consistent code, we can see that, due to the extraction of the stellar disk, the $\ell= 2$ is the only mode that is ``excited''.}
\vspace{12pt}
\label{fig:appendix_powerspectrum_norm_RELAX}
\end{figure*}

The run in isolation was performed at HR, since we wanted to test if our models are stable against numerical noise. In this run, we use our MW model from Table \ref{tab:ICfiles} and evolve the galaxy in isolation for $7.7$~Gyr, which is exactly as long as our simulation runs. In Figure~\ref{fig:appendix_powerspectrum_norm_RELAX}, we show the power spectrum as in Figure~\ref{fig:powerspectrum_norm}, in which we can see that, except for the $\ell = 2$ mode (which corresponds to the extracted stellar disk), we do not excite any DM modes throughout the entire simulation.

\end{document}